\documentclass[namedreferences]{solarphysics}

\usepackage{amsmath}
\usepackage[optionalrh,showbiblabels]{spr-sola-addons} 
\usepackage{graphicx}        
\usepackage{xcolor}           

\renewcommand{\vec}[1]{{\mathbfit #1}}

\newcommand{\aap}{    {\it Astron. Astrophys.}}

\newcommand{\apj}{    {\it Astrophys. J.}}
\newcommand{\apjl}{   {\it Astrophys. J. Lett.}}

\newcommand{\mnras}{  {\it Mon. Not. Roy. Astron. Soc.}}

\newcommand{\solphys}{{\it Solar Phys.}}
 
\newcommand{\ssr}{    {\it Space Sci. Rev.}} 
\chardef\us=`\_

\begin{document}

\begin{article}
\begin{opening}

\title{Effect of Thermal Conductivity, Compressive Viscosity and Radiative Cooling on the Phase Shift of Propagating Slow Waves with and without Heating--Cooling Imbalance}

\author[addressref=aff1]{\inits{Abhinav}\fnm{Abhinav}~\lnm{Prasad}}
\author[addressref={aff1},corref,email={asrivastava.app@itbhu.ac.in}]{\inits{A.K.}\fnm{A.K.}~\lnm{Srivastava}}
\author[addressref=aff2]{\inits{Tongjiang}\fnm{Tongjiang}~\lnm{Wang}}

\address[id=aff1]{Department of Physics, Indian Institute of Technology (BHU), Varanasi-221005, UP, India.}
\address[id=aff2]{The Catholic University of American and NASA Goddard Space Flight Center, Code 671, Greenbelt, MD, 20771, USA.}

\runningauthor{A. Prasad et al.}
\runningtitle{Phase Shift of Slow Waves in Solar Corona}

\begin{abstract}
 We study the phase shifts of propagating slow  magnetoacoustic waves in solar coronal loops invoking the effects of thermal conductivity,  compressive viscosity, radiative losses, and heating--cooling imbalance. We derive the general dispersion relation and solve it to determine the phase shifts of density and temperature perturbations relative to the velocity and their dependence on the equilibrium parameters of the plasma such as the background density [$\rho_0$] and temperature [$T_0$]. We estimate the phase difference [$\Delta \phi$] between density and temperature perturbations and its dependence on $\rho_0$ and $T_0$. The role of radiative losses, along with the heating--cooling imbalance for an assumed specific heating function [$H(\rho, T) \propto \rho^{-0.5} T^{-3}$] in the estimation of the phase shifts is found to be significant for the high-density and low-temperature loops. Heating--cooling imbalance can significantly increase the phase difference ($\Delta \phi \approx 140^\circ$) for the low-temperature loops compared to the constant-heating case ($\Delta \phi \approx 30^\circ$). We derive a general expression for the polytropic index [$\gamma_{\rm eff}$] using the linear MHD model. We find that in the presence of thermal conduction alone, $\gamma_{\rm eff}$ remains close to its classical value $5/3$ for all the considered $\rho_0$ and $T_0$ observed in typical coronal loops. We find that the inclusion of radiative losses (with or without heating--cooling imbalance) cannot explain the observed polytropic index under the considered heating and cooling models. To make the expected $\gamma_{\rm eff}$ match the observed value of $1.1 \pm 0.02$ in typical coronal loops, the thermal conductivity needs to be enhanced by an order of magnitude compared to the classical value. However, this conclusion is based on the presented model and needs to be confirmed further by considering more realistic radiative functions. We also explore the role of different heating functions for typical coronal parameters and find that although the $\gamma_{\rm eff}$ remains close to $5/3$, but the phase difference is highly dependent on the form of the heating function.
 
\end{abstract}
\end{opening}

\section{Introduction}
{
Propagating disturbances and/or brightenings are observed in a variety of coronal magnetic structures, e.g. polar plumes in coronal holes \citep[e.g.][and references therein]{1997ApJ...491L.111O,2000ApJ...529..592O,1998ApJ...501L.217D,2011A&A...528L...4K,2014ApJ...793..117S,2020ApJ...900L..19C}, and coronal fan loops in active regions \citep[e.g.][and references therein]{2000A&A...355L..23D,2001A&A...370..591R,2002A&A...387L..13D,2002SoPh..209...89D,2002SoPh..209...61D,2003A&A...404L..37M,2009ApJ...696.1448W,2009A&A...503L..25W,2013ApJ...778...26U,2015RAA....15.1832M}. The consensus view is that these bright disturbances and associated field-aligned propagation are propagating slow magnetoacoustic waves in the solar atmosphere, which also  exhibit a strong dissipative physical nature \citep[e.g.][and references cited there]{1999ApJ...514..441O,2000ApJ...533.1071O,2003A&A...408..755D,2004A&A...415..705D,2004A&A...425..741D}. There is a large body of literature in solar physics that reports the intensity and/or velocity oscillations present in the quiet Sun, above sunspots in active regions, and in coronal holes, revealing the ubiquitous presence of slow magnetoacoustic waves \citep[e.g.][and references cited there]{2001A&A...370..591R,2002A&A...387..642O,2003A&A...404L..37M,2005ApJ...624L..61D,2008A&A...481L..95S,2003A&A...404L...1K,2018MNRAS.479.5512K,2020A&A...634A..63K}. On many occasions in recent years, observations related to the intensity fluctuations in coronal structures have given rise to debates on whether they are slow waves or periodic plasma flows  \citep[e.g.][and references therein]{2010ApJ...722.1013D,2011ApJ...727L..37T,2012APC...456..P91,2012ApJ...754..111O,2015SoPh..290..399D,2016A&A...592L...8M,2016GU...216...395}.

As far as propagating slow waves are concerned, \citet{2004ApJ...616.1232K} have demonstrated that the intensity perturbations of these  waves depend upon a variety of physical conditions and parameters of the localized solar corona, e.g. dissipation of the wave energy via thermal conduction, pressure and temperature gradients, and mutual action between wave propagation and the plasma flows. Over the last two decades there have been numerous attempts to model these intensity oscillations in the form of slow waves in the gravitationally stratified non-ideal solar atmosphere with realistic temperature and magnetic field conditions, in typical corona where thermal conductivity, viscosity, and radiative losses were the dominant dissipation mechanisms \citep[e.g.][and references cited there]{1999ApJ...514..441O,2000ApJ...533.1071O,2000A&A...362.1151N,2001A&A...379.1106T,2003A&A...408..755D,2004A&A...415..705D,2004A&A...425..741D}. With the advent of high-resolution and multi-channel spectrometers/spectrographs -- e.g. {\it Hinode/EUV Imaging Spectrometer} (EIS), \textit{Interface Region Imaging Spectrograph} (IRIS) -- and imagers -- e.g. \textit{Solar Dynamics Observatory} (SDO)/\textit{Atmospheric Imaging Assembly} (AIA), {\it Hinode}) -- forward modelling became a prominent tool to further understand the slow-wave observations in the solar atmosphere (e.g. intensity oscillations) by modelling and estimating the wave variables \citep{2009A&A...494..339O,2015ApJ...813...33F,2016ApJ...828...72M}. \citet{2009A&A...494..339O} have studied the effects of thermal conduction on the phase relationships between the velocity and the thermal energy and density perturbations for the slow waves. These properties measured from observations could be important for understanding the dissipative agents, e.g. thermal conduction, and plasma properties of the medium leading this phase shift, and also help differentiating the wave dynamics from the normal field-aligned plasma flows. \citet{2018ApJ...868..149K} have analysed the oscillation amplitude in temperature (thus thermal energy) and density (thus intensity) corresponding to slow magnetoacoustic waves, and estimated the polytropic index in the solar corona. They have found that thermal conduction tends to be suppressed in hotter loops while it is higher in relatively cool loops. \citet{2011ApJ...727L..32V} have used  the time-dependent wave signals from multiple spectral lines of {\it Hinode}/EIS formed at a range of temperatures, and they deduced the relationship between relative density and temperature perturbations and estimated the effective adiabatic index (or polytropic index) to be $\approx$1.10,  from which they have conjectured that thermal conduction along the magnetic field is very efficient in the solar corona, based on the diagnosis by coronal seismology of slow magnetoacoustic waves. \citet{2015ApJ...811L..13W} have derived the time-dependent temperature and electron-density variations using six AIA extreme-ultraviolet filters of SDO. They find that these temporal variations are nearly in phase. The estimated polytropic index from the temperature and density perturbations is found to be $\approx$1.64, which approximately matches the adiabatic index of $5/3$ for an ideal monoatomic gas. It was derived by \citet{2015ApJ...811L..13W} that the thermal conductivity is suppressed by at least a factor of three in the hot flare loop at 9$\,\,$MK and above, which is supported by the later report of \citet{2018ApJ...868..149K}. Moreover, the viscosity coefficient was determined by \citet{2015ApJ...811L..13W} using coronal seismology from the observed  slow-mode waves when they considered compressive viscosity  as the only dissipation mechanism. They found that the interpretations of the observed rapid damping required the classical compressive viscosity coefficient to be enhanced by more than an order of magnitude. \citet{2019ApJ...886....2W} further improved this method based on a parametric study of a nonlinear 1D MHD model including both thermal conduction and compressive viscosity. In conclusion, slow waves are strongly influenced by the dissipative agents, e.g. thermal conductivity, and local plasma conditions. Therefore, the observed phase shifts of intensity (or density) and temperature perturbations and their dependence on the plasma parameters in slow waves can provide a unique diagnostic capability of the localized solar atmosphere where they propagate and dissipate. However, the previous studies suggest that such phase shifts are highly sensitive to the physical conditions of the solar atmosphere, so it requires realistic modelling to infer them correctly in the presence of  relevant plasma processes.

Different heating and cooling mechanisms in coronal loops simultaneously work on establishing the propagation and dissipation of both propagating and standing slow waves, and thermal conduction is found to be the dominant dissipative agent for the slow waves in normal coronal loops \citep[see reviews by][and references therein]{2009SSRv..149...65D,2011SSRv..158..397W,2021SSRv...accep}. Moreover, the specific heating mechanism is still unknown in the corona, therefore many previous theoretical models of MHD waves have considered heating functions depending upon the physical parameters, e.g. temperature, density, and magnetic field, which are static or time-dependent, e.g. due to wave-induced heating--cooling imbalance \citep[e.g.][and references therein]{2019A&A...628A.133K,2020A&A...644A..33K,2020SSRv..216..140V,2020arXiv201114519P}. \citet{2020arXiv201114519P} studied the damping of standing slow MHD oscillations for a wide range of coronal loops by assuming a specific heating function [$H(\rho,T) \propto \rho^{-0.5}T^{-3}$] and explained that the inclusion of heating--cooling imbalance can better account for the scaling relationship between observed periods and damping times of \textit{Solar Ultraviolet Measurements of Emitted Radiation} (SUMER) oscillations. In their work, this choice of heating function was based on getting a good oscillation quality factor (i.e. ratio of decay time to period) on the order of one to two consistent with the SUMER observations for the standing waves, and it also lies in the regime of enhanced damping as defined by \citet{2019A&A...628A.133K}. In the present study we explore the effect of the same heating function on the phase shifts of propagating slow waves. As mentioned above, while thermal conduction is known as a well-established mechanism in determining even the phases of temperature (or thermal energy) and density perturbations of the medium due to slow waves, it is obvious that the additional effects of heating--cooling imbalance along with other dissipative mechanisms must influence this important physical property. \citet{2009A&A...494..339O} have studied such a phase relationship between thermal energy and density perturbation of propagating slow waves due to thermal conduction only. \citet{2011ApJ...727L..32V} have derived the phase relation between these physical quantities using the energy equation only.

In the present article, we derive a new comprehensive theory of propagating slow waves taking into consideration all of the dissipative effects (including thermal conduction, compressive viscosity, and radiation), along with the heating--cooling imbalance. We also invoke a constant heating rate and compare the results with the case of heating--cooling imbalance. By defining the viscous, thermal, radiative, and heating--cooling imbalance ratios to characterize the different dissipations, we derive a new dispersion relation using a new theoretical model incorporating the effects of thermal conductivity, viscosity, radiative losses, and heating--cooling imbalance, and we study the phase relationships between different physical parameters for the propagating slow waves in the solar coronal loops based on this dispersion relation. By numerical and analytical analysis, we determine the phase shifts of perturbed density and temperature relative to the velocity and their dependence on the loop background density and temperature. We have also estimated the phase difference between density and temperature. We derive the generalized expression for the polytropic index, and we study its dependence on the physical parameters using our model in order to  understand recent measurements. Section 2 elucidates the entire model in a very detailed manner describing each step and computed parameters. The theoretical results are described in various sub-sections of Section 3. The last section presents the discussion and conclusions, and it outlines the most important new science dealt with by the current model, and also its future prospective.
}
\section{Theoretical Model}
{
We consider the basic MHD equations as below \citep{1965RvPP....1..205B,2014masu.book.....P}\newline The mass conservation equation is
\begin{equation}
    \frac{\partial \rho}{\partial t} + \nabla \cdot (\rho {\vec{v}}) = 0
\end{equation}
The momentum conservation equation is
\begin{equation}
    \rho \left(\frac{\partial \vec{v}}{\partial t} + \vec{v}\cdot \nabla \vec{v}\right) = -\nabla p + \frac{1}{\mu_0}(\nabla \times \vec{B})\times \vec{B} - \nabla \cdot \mathbf{\Pi}
\end{equation}
The energy conservation equation is
\begin{equation}
     \frac{\partial \epsilon}{\partial t} + \vec{v} \cdot \nabla \epsilon  = -\frac{p}{\rho}\nabla \cdot \vec{v} + \frac{1}{\rho}(\nabla \cdot \boldsymbol{\kappa})\cdot \nabla T -  \chi\rho T^\alpha + H(\rho,T) + \frac{Q_{\rm vis}}{\rho}
\end{equation}
The ideal gas equation is
\begin{equation}
    \rho = \frac{mp}{\mathrm{k}_{\rm B}T}
\end{equation}
Here $\rho$, $\vec{v}$, $p$, $\vec{B}$, and $T$ are the mass density, velocity field, pressure, magnetic field, and temperature. $\mathrm{k}_{\rm B}$ is the Boltzmann constant and $m$ is the mean particle mass equal to $0.6\,{\text m}_{\rm p}$ where ${\text m}_{\rm p}$ is the proton mass. $\mathbf{\Pi}$ is the viscous tensor, $\boldsymbol{\kappa}$ is the thermal conductivity tensor, and $Q_{\rm vis}$ represents the viscous heating. The radiative losses are represented by the function $\chi\rho T^\alpha$ where $\chi$ and $\alpha$ are constants for the solar corona. We consider $\alpha = -0.5$ and $\chi = \frac{10^{-32}}{m^2}$, which is a good approximation for analytical modelling valid for coronal abundances and for a temperature range from $2 \times 10^5$\,K $< T_0 < 10^7$\,K \citep{2014masu.book.....P}. A density and temperature-dependent heating function [$H(\rho,T) = h\rho^a T^b$] is assumed as considered by \citet{2019A&A...628A.133K} where $a$ and $b$ are free parameters. The gravitational effects are ignored throughout the analysis.

In order to model the propagating slow waves in the solar corona, we simplify the above MHD equations under a set of approximations. The viscous tensor is assumed to be highly anisotropic with compressive viscosity dominating in comparison to shear viscosity. Thermal conductivity across the field line is ignored in comparison to the large thermal conductivity coefficient along the magnetic-field lines. We also consider the infinite magnetic-field approximation where all of the perturbations are directed along the stiff magnetic-field lines and the plasma-$\beta$ is zero. In the present analysis we consider the magnetic-field lines to be homogeneous and along the $z$-direction. This approximation greatly simplifies the general MHD equations by converting them into their 1D analogues along the magnetic-field line, which is useful for the study of slow waves \citep[and references therein]{2003A&A...408..755D,2019A&A...628A.133K}. The convective derivative is simplified as below under the 1D approximation 
\begin{equation}
    \frac{\partial}{\partial t} + \vec{v}\cdot \nabla  \approx  \frac{\partial}{\partial t} + v\frac{\partial}{\partial z}
\end{equation}
Here $v$ is the $z$-component of velocity field [$\vec{v}$]\newline
The viscous and thermal conductivity tensor are simplified in 1D as below \citep{1965RvPP....1..205B}:
\begin{equation}
    \nabla \cdot \mathbf{\Pi} \approx -\frac{4\nu}{3}\frac{\partial^2v}{\partial z^2} \hat{\mathit{\mathbf{e}}}_z \,\,\,\,\,\,\,\, (\text{viscous force}),
\end{equation} 
\begin{equation}
    Q_{\rm vis} = \frac{\nu}{3}\left( \nabla \cdot \vec{v} - 3\frac{\partial v}{\partial z}\right)^2 \approx \frac{4\nu}{3}\left( \frac{\partial v}{\partial z} \right)^2 \,\,\,\, (\text{viscous heating}),
\end{equation}
\begin{equation}
    (\nabla \cdot \boldsymbol{\kappa}) \cdot \nabla T \approx \frac{\partial}{\partial z}\left(\kappa \frac{\partial T}{\partial z}\right)\,\,\,\,\,\, (\text{thermal heating}).
\end{equation}
The coefficients of thermal conductivity and viscosity along the field line are taken as below \citep{1965RvPP....1..205B}:
\begin{equation*}
    \kappa = \kappa_0 T^{5/2},
\end{equation*}
\begin{equation*}
    \nu = \nu_0 T^{5/2},
\end{equation*}
where $\kappa_0 = 9 \times 10^{-12}$ W ${\text m}^{-1}{\text K}^{-1}$ and $\nu_0 = 10^{-17}$\,kg\,m$^{-1}$ ${\text s}^{-1}$\newline
Thus we finally have the following simplified MHD equations as below. Also keep in mind that under the 1D approximation all of the variables are only a function of $z$ and $t$.\newline
The mass conservation equation is
\begin{equation}
    \frac{\partial \rho}{\partial t} = -\rho \frac{\partial v}{\partial z} - v \frac{\partial \rho}{\partial z}.
\end{equation}
The momentum conservation equation is
\begin{equation}
    \frac{\partial v}{\partial t} = -v\frac{\partial v}{\partial z} - \frac{1}{\rho}\frac{\partial p}{\partial z} + \frac{4\nu}{3\rho}\frac{\partial^2v}{\partial z^2}.
\end{equation}
Note that the above equation is a simplified 1D $z$-component of Equation (2) and here the Lorentz force along the field line is zero due to the assumption of a homogeneous magnetic field.\newline
The energy conservation equation is
\begin{equation}
    \frac{\partial \epsilon}{\partial t} = -v\frac{\partial \epsilon}{\partial z} - \frac{p}{\rho}\frac{\partial v}{\partial z} + \frac{1}{\rho}\frac{\partial}{\partial z} \left( \kappa \frac{\partial T}{\partial z} \right) - \chi\rho T^\alpha + h\rho^a T^b + \frac{4\nu}{3\rho}\left( \frac{\partial v}{\partial z} \right)^2.
\end{equation}
The initial equilibrium is set up by balancing the radiative losses and unknown coronal heating:
\begin{equation}
    \chi \rho_0 T_0^\alpha = H(\rho_0, T_0) = h\rho_0^a T_0^b
\end{equation}
Thus the coefficient $h$ is calculated based on the initial equilibrium conditions of the plasma,\newline
\begin{equation*}
    h = \chi \rho_0^{1-a}T_0^{\alpha-b}.
\end{equation*}
The ideal gas equation is
\begin{equation}
    \rho = \frac{mp}{{\text k}_{\rm B}T}.
\end{equation}
The energy density is given by
\begin{equation}
    \epsilon = \frac{p}{(\gamma-1)\rho} = \frac{{\text k}_{\rm B}T}{m(\gamma-1)}.
\end{equation}
From Equation 14 we can re-write Equation 11 by eliminating $\epsilon$:
\begin{multline}
    \frac{\partial T}{\partial t} = -v\frac{\partial T}{\partial z} - (\gamma-1)T\frac{\partial v}{\partial z} + \frac{m(\gamma-1)\kappa_0}{\rho\, {\text k}_{\rm B}}\frac{\partial}{\partial z} \left( T^{5/2} \frac{\partial T}{\partial z} \right)\\ - \frac{m\chi(\gamma-1)}{{\text k}_{\rm B}}  \rho T^\alpha + \frac{mh(\gamma-1)}{{\text k}_{\rm B}}  \rho^a T^b + \frac{m(\gamma-1)}{{\text k}_{\rm B}}\frac{4\nu_0T^{5/2}}{3\rho}\left( \frac{\partial v}{\partial z} \right)^2.
\end{multline}
\subsection{Dimensionless Equations}
{
For further analysis we make the 1D MHD equations dimensionless with respect to the equilibrium parameters \citep{2003A&A...408..755D}. The equilibrium plasma is assumed to be homogeneous and we non-dimensionalize the equations with respect to the constant background density [$\rho_0$], pressure [$p_0$], and temperature [$T_0$]. We also use the timescale $\tau \approx 300$ seconds for non-dimensionalization, which is the wave period of propagating slow waves \citep{2009A&A...494..339O}. Note that $v$, $\rho$, $T$, and $p$ in the following equations are dimensionless. \newline
The mass equation is
\begin{equation}
    \frac{\partial \rho}{\partial t} = -\rho \frac{\partial v}{\partial z} - v \frac{\partial \rho}{\partial z}.
\end{equation}
The momentum equation is
\begin{equation}
    \frac{\partial v}{\partial t} = -v\frac{\partial v}{\partial z} - \frac{p_{0}}{c_{s0}^2 \rho_{0}}\frac{1}{\rho}\frac{\partial p}{\partial z} + \frac{eT^{5/2}}{\rho}\frac{\partial^2v}{\partial z^2},
\end{equation}
since $c_{s0}^2 = \frac{\gamma p_{0}}{\rho_{0}}$, so we can rewrite the above equation as
\begin{equation}
     \frac{\partial v}{\partial t} = -v\frac{\partial v}{\partial z} - \frac{1}{\gamma \rho}\frac{\partial p}{\partial z} + \frac{eT^{5/2}}{\rho}\frac{\partial^2v}{\partial z^2}.
\end{equation}
The energy equation is
\begin{multline}
    \frac{\partial T}{\partial t} = -v\frac{\partial T}{\partial z} - (\gamma-1)T\frac{\partial v}{\partial z} + \frac{\gamma d}{\rho} \frac{\partial}{\partial z} \left( T^{5/2} \frac{\partial T}{\partial z} \right) - \gamma r \rho T^\alpha + \gamma r \rho^a T^b +\\ +\frac{m(\gamma-1)}{{\text k}_{\rm B}}\frac{4\nu_0 T_{0}^{3/2}}{3\rho_{0}\tau}\frac{T^{5/2}}{\rho}\left( \frac{\partial v}{\partial z} \right)^2.
\end{multline}
The ideal gas equation is
\begin{equation}
    \rho = \frac{p}{T},
\end{equation}
where we have the dimensionless ratios as defined by \citet{2003A&A...408..755D}:
\begin{equation}
    e = \frac{4\nu_0 T_{0}^{5/2}}{3\rho_{0}\tau c_{s0}^2}\,\,\, \text{(Viscous ratio)}
\end{equation}
\begin{equation}
    d =\frac{(\gamma-1)\kappa_0T_{0}^{7/2}\rho_0}{\tau \gamma^2 p_0^2}\,\,\, \text{(Thermal ratio)}
\end{equation}
\begin{equation}
    r = \frac{(\gamma-1)\tau \rho_{0}^2}{\gamma p_0} \chi T_{0}^{\alpha}\,\,\, \text{(Radiative  ratio)}
\end{equation}
The density, temperature, and pressure in equilibrium satisfy the following relation
\begin{equation}
    \rho_{0} = \frac{mp_{0}}{{\text k}_{\rm B}T_{0}}.
\end{equation}
}
\subsection{Linearization}
{
We consider first-order perturbations to the equilibrium as below:
\begin{equation}
    \rho = 1 + \rho_1,
\end{equation}
\begin{equation}
    T = 1 + T_1,
\end{equation}
\begin{equation}
    v = v_1,
\end{equation}
\begin{equation}
    p = 1 + p_1.
\end{equation}
Further we write the linearized MHD equations as below by substituting the above quantities into Equations 16\,--\,20. The linearization is done by considering a constant and homogeneous background plasma. 
\begin{equation}
    \frac{\partial \rho_1}{\partial t} = -\frac{\partial v_1}{\partial z},
\end{equation}
\begin{equation}
    \frac{\partial v_1}{\partial t} = -\frac{1}{\gamma}\frac{\partial p_1}{\partial z} + e\frac{\partial^2 v_1}{\partial z^2},
\end{equation}
\begin{equation}
    \frac{\partial T_1}{\partial t} = -(\gamma-1)\frac{\partial v_1}{\partial z} + \gamma d  \left(  \frac{\partial^2 T_1}{\partial z^2} \right) - \gamma r(\alpha T_1 + \rho_1) + \gamma r(bT_1 + a\rho_1),
\end{equation}
\begin{equation}
    p_1=  \rho_1 +  T_1.
\end{equation}
}
\subsection{Fourier Decomposition}
{
Considering Fourier decomposition of the form
\begin{equation}
    f = \hat{f}{\text e}^{{\text i}(kz-\omega t)}
\end{equation}
and substituting it in Equations 29\,--\,32 we get
\begin{equation}
    \omega \hat{\rho}_1 - k \hat{v}_1 = 0,
\end{equation}
\begin{equation}
    -{\text i}\gamma \omega\hat{v}_1 = -{\text i}k\hat{p}_1 - e\gamma k^2\hat{v}_1,
\end{equation}
\begin{multline}
    -{\text i}\omega \hat{T}_1 = -{\text i}k(\gamma-1)\hat{v}_1 - \gamma dk^2\hat{T}_1 - \gamma r(\alpha \hat{T}_1 + \hat{\rho}_1) + \gamma r(b\hat{T}_1 + a\hat{\rho}_1),
\end{multline}
\begin{equation}
    \hat{p}_1=  \hat{\rho}_1 +  \hat{T}_1.
\end{equation}
Simplifying the above equations we have 
\begin{equation}
    \omega \hat{\rho}_1 - k \hat{v}_1 = 0,
\end{equation}
\begin{equation}
    {\text i}k\hat{\rho}_1 + \gamma(ek^2 - {\text i}\omega )\hat{v}_1 + {\text i}k\hat{T}_1 = 0,
\end{equation}
\begin{multline}
    \gamma r(a-1)\hat{\rho}_1 - {\text i}k(\gamma-1)\hat{v}_1 + ({\text i}\omega +\gamma r(b-\alpha) - \gamma dk^2)\hat{T}_1 = 0.
\end{multline}
We write the Equations 38\,--\,40 in matrix form:
\begin{multline}
    \begin{bmatrix}
       \omega & -k & 0   \\
       {\text i}k & \gamma(ek^2 - {\text i}\omega) & {\text i}k \\
        \gamma r(a-1) & - {\text i}k(\gamma-1) & {\text i}\omega +\gamma r(b-\alpha) - \gamma dk^2  \\
    \end{bmatrix}
    \begin{bmatrix}
       \hat{\rho}_1\\
       \hat{v}_1\\
       \hat{T}_1
    \end{bmatrix}
    = M 
    \begin{bmatrix}
       \hat{\rho}_1\\
       \hat{v}_1\\
       \hat{T}_1
    \end{bmatrix}
    =0
\end{multline} 
For a non-trivial solution we make the Det($M$) to be zero:
\begin{gather}
{\text D}{\text e}{\text t}(M)=
    \begin{vmatrix}
        \omega & -k & 0   \\
       {\text i}k & \gamma(ek^2 - {\text i}\omega) & {\text i}k \\
        \gamma r(a-1) & - {\text i}k(\gamma-1) & {\text i}\omega +\gamma r(b-\alpha) - \gamma dk^2  \\
    \end{vmatrix}.
\end{gather}
Solving the above determinant for propagating waves, i.e. where $k$ is complex and $\omega$ is real, we get the following expression
\begin{equation}
    Ak^4 + Bk^2 + C=0,
\end{equation}
where 
\begin{equation}
    A = -\gamma d({\text i} + \gamma \omega e),
\end{equation}
\begin{multline}
    B = ({\text i} + \gamma \omega e)({\text i}\omega + \gamma r(b - \alpha)) + {\text i}\gamma^2\omega^2d - \omega(\gamma-1) - {\text i}\gamma r(a-1),
\end{multline}
\begin{equation}
    C = -{\text i}\gamma \omega^2({\text i}\omega + \gamma r(b - \alpha)).
\end{equation}
For propagating waves the wavenumber $k$ is a complex number, i.e. \newline $k = k_{\rm r} + {\text i}k_{\rm i} = k_m {\text e}^{{\text i}\phi}$, where $k_{\rm m} = \sqrt{k_{\rm r}^2 + k_{\rm i}^2}$, while $\omega$ is real, \newline
since we have
\begin{equation}
    v_1 = \hat{v}_1 {\text e}^{{\text i}(k_{\rm r}z-\omega t)}{\text e}^{-k_{\rm i}z}.
\end{equation}
Thus from Equation 38 we write
\begin{equation}
    \rho_1 = \frac{k_{\rm m}\hat{v}_1}{\omega}{\text e}^{{\text i}(k_{\rm r}z+\phi-\omega t)}{\text e}^{-k_{\rm i}z}.
\end{equation}
The phase shift of $\rho_1$ with respect to $v_1$ is given by
\begin{equation}
    z_{\rho} = \frac{\phi}{k_{\rm r}} = \frac{1}{k_{\rm r}}\tan^{-1}\left( \frac{k_{\rm i}}{k_{\rm r}} \right).
\end{equation}
For convenience, hereafter we will refer to $z_\rho$ as the density phase shift \newline
From Equation 39 we have
\begin{equation}
    \hat{T}_1 = \left(\frac{\gamma \omega }{k} + {\text i}\gamma e k \right)\hat{v}_1 - \hat{\rho}_1 = \hat{v}_1(\alpha_1 {\text e}^{-{\text i}\phi} + {\text i}\beta_1 {\text e}^{{\text i}\phi} - \beta {\text e}^{{\text i}\phi}),
\end{equation}
where 
\begin{equation}
    \alpha_1 = \frac{\gamma \omega }{k_{\rm m}} \,\,\,,\,\,\,\, \beta = \frac{ k_{\rm m}}{\omega}\,\,\,,\,\,\,\, \beta_1 = \gamma e k_{\rm m}.
\end{equation}
Writing $T_1$ as 
\begin{equation}
    T_1 = \hat{v}_1R{\text e}^{{\text i}(k_{\rm r}z+\Phi-\omega t)}{\text e}^{-k_{\rm i}z},
\end{equation}
where 
\begin{equation}
    R = \sqrt{((\alpha_1 - \beta)\cos\phi- \beta_1\sin\phi)^2 + (\beta_1\cos\phi -(\alpha_1 + \beta)\sin\phi)^2},
\end{equation}
\begin{equation}
    \Phi = \tan^{-1}\left( \frac{\beta_1\cos\phi -(\alpha_1 + \beta)\sin\phi}{(\alpha_1 - \beta)\cos\phi- \beta_1\sin\phi} \right).
\end{equation}
The phase shift of $T_1$ with respect to $v_1$ is thus given as
\begin{equation}
    z_T = \frac{\Phi}{k_{\rm r}},
\end{equation}
and hereafter we will refer to $z_T$ as the temperature phase shift.
}
}
\section{Theoretical Results}
We study the effects of thermal conductivity, viscosity, radiative losses, and heating--cooling imbalance on the density and temperature phase shifts in the subsequent sections. In Section 3.1 we consider the effect of only thermal conductivity and compare our results with some of the previous work by \citet{2009A&A...494..339O}. We systematically add the viscous dissipation, radiative losses, and heating--cooling imbalance into our model and discuss their influence on the phase shift for a range of coronal loops under Sections 3.2, 3.3 and 3.4 respectively. We consider a range of densities ($\approx$ 2 $\times 10^{-13}$\,--\,4$\times$10$^{-12}$\,kg\,m$^{-3}$) and a range of temperatures ( 0.1\,--\,2.1\,MK). In Section 3.5, we derive a general expression for the polytropic index and use it to discuss the observed data given by some of the previous works of \citet{2018ApJ....860...107}, \citet{2018ApJ...868..149K}, and \citet{2011ApJ...727L..32V}. Finally, we provide an estimate of the thermal [$d$] and radiative ratio [$r$] that can explain the observed polytropic index.

\subsection{Analysis of the Effect of Thermal Conductivity}
{
Considering only the effect of thermal conductivity ($e=r=0$) we have the simplified expression
\begin{equation}
    Ak^4 + Bk^2 + C=0,
\end{equation}
where 
\begin{equation}
    A = -{\text i}d,
\end{equation}
\begin{equation}
    B = -\omega + {\text i} d \omega^2\gamma,
\end{equation}
\begin{equation}
    C = \omega^3. 
\end{equation}
The phase difference between temperature and density perturbation can be written as 
\begin{equation}
    \Delta z = z_{\rho} - z_{T} = \frac{\Delta \phi}{k_r}.
\end{equation}
For ease we will simply refer to $\Delta \phi$ as the phase difference from now on.\newline
We considered
\begin{equation}
    \rho_1 = \hat{\rho}_1{\text e}^{{\text i}(kz-\omega t)},
\end{equation}
\begin{equation}
    T_1 = \hat{T}_1 {\text e}^{{\text i}(kz-\omega t-\Delta \phi)},
\end{equation}
we substitute the above solutions in the energy equation, which is simplified as following when only thermal conductivity is considered:
\begin{equation}
    \frac{\partial T_1}{\partial t} = -(\gamma-1)\frac{\partial v_1}{\partial z} + \gamma d  \left(  \frac{\partial^2 T_1}{\partial z^2} \right),
\end{equation}
\begin{equation}
     \frac{\partial T_1}{\partial t} = (\gamma-1)\frac{\partial \rho_1}{\partial t} + \gamma d  \left(  \frac{\partial^2 T_1}{\partial z^2} \right).
\end{equation}
We thus have
\begin{equation}
    {\text e}^{-{\text i}\Delta \phi}(\gamma d k^2 - {\text i}\omega)\hat{T}_1 = -{\text i}\omega (\gamma-1)\hat{\rho}_1,
\end{equation}
\begin{equation}
    \left( \frac{{\text i}\gamma d }{\omega}(k_{\rm r}^2 - k_{\rm i}^2) + \frac{1}{\omega}(\omega - 2\gamma d k_{\rm r} k_{\rm i})  \right)(\cos \Delta \phi - {\text i}\sin \Delta \phi)\hat{T}_1 = (\gamma-1)\hat{\rho}_1.
\end{equation}
Taking the imaginary component of the above equation, we can write
\begin{equation}
    \tan \Delta \phi = \frac{\gamma d (k_{\rm r}^2 - k_{\rm i}^2)}{\omega - 2\gamma d k_{\rm r} k_{\rm i}},
\end{equation}
and under the weak-damping assumption $k_\mathrm{i} \approx 0$ and $k_\mathrm{r} \approx \omega $ we get
\begin{equation}
    \tan \Delta \phi = \omega \gamma d \implies \Delta \phi \approx \omega \gamma d.
\end{equation}
This expression is consistent with and the same as Equation 12 of \citet{2018ApJ....860...107}.\newline
Considering the weak thermal conduction approximation ($d \ll 1$), the phase shift between $\rho_1$ and $T_1$ can be estimated as by \citet{2009A&A...494..339O}, assuming
\begin{equation}
    k = \omega + {\text i}d\Omega.
\end{equation}
Ignoring terms of order $d^2$ and substituting in Equation 56 we get
\begin{equation}
    \Omega = \frac{(\gamma-1)\omega^2 }{2}.
\end{equation}
With Equations 49 and 55 we can write
\begin{equation}
    z_{\rho} = \frac{(\gamma-1)d}{2},
\end{equation}
\begin{equation}
    z_{T} = \frac{-(\gamma+1)d}{2},
\end{equation}
from Equation 60 and considering $k_{\rm r} = \omega $ we can write
\begin{equation}
    \Delta \phi = k_{\rm r}(z_{\rho} - z_T) = \omega \gamma d.
\end{equation}
This expression is the same as Equation 68, which is expected since $d \ll 1$ implies the weak-damping assumption by thermal conduction.\newline
We solve Equation 56 for $k$ using the Wolfram {\it Mathematica} environment from 2016. Note that we consider $\omega = 2\pi$ throughout our analysis.
We calculate the temperature and density phase shifts by substituting the obtained solution of $k$ into Equations 49 and 55. Our objective is to understand the variation of phase shifts over a range of background densities and temperatures of coronal loops. In order to facilitate our further calculations and analysis we define dimensionless equilibrium density and temperature as
\begin{equation}
    \rho^{\prime} = \frac{\rho_0}{\rho_{00}},
\end{equation}
and
\begin{equation}
    T^{\prime} = \frac{T_0}{T_{00}},
\end{equation}
where
\begin{equation*}
    \rho_{00} = 1.67 \times 10^{-12} \,\,\text{kg}\, \text{m}^{-3},
\end{equation*}
\begin{equation*}
    T_{00} = 1\,{\text M}{\text K}.
\end{equation*}
Thus we have
\begin{equation}
    d = \frac{(\gamma-1)\kappa_0 m^2}{\tau \gamma^2 {\text k}_{\rm B}^2} \frac{T_{00}^{3/2}}{\rho_{00}}\frac{(T^{\prime})^{3/2}}{\rho^{\prime}}
\end{equation}
\begin{equation}
    e = \frac{4\nu_0 m}{3\gamma \tau {\text k}_{\rm B}} \frac{T_{00}^{3/2}}{\rho_{00}}\frac{(T^{\prime})^{3/2}}{\rho^{\prime}}
\end{equation}
\begin{equation}
    r = \frac{(\gamma-1)\tau \chi m}{\gamma {\text k}_{\rm B}} \rho_{00}T_{00}^{\alpha-1} (\rho^{\prime}(T^{\prime})^{\alpha-1})
\end{equation}
In Table 1 we have summarized the values of dimensionless ratios calculated from Equations 76\,--\,78 over the range of equilibrium densities and temperatures considered throughout our study.
\begin{table}[t!]
\begin{tabular}{ p{2cm} p{1.65cm} p{1.65cm} p{1.8cm} }

\hline
{\bf Ratios} & { \bf Thermal ratio [$d$]} & {\bf Viscous \newline ratio [$e$]} & {\bf Radiative ratio [$r$]}\\
\hline
 $T_0$ = 0.1 $T_{00}$ \newline $\rho_0 = \rho_{00}$  & 0.0007& 0.000036 & 4.577 \\
 \hline
  $T_0$ = 2.1 $T_{00}$ \newline $\rho_0 = \rho_{00}$ & 0.069 & 0.0035 & 0.047\\
\hline
  $T_0$ = $T_{00}$ \newline $\rho_0 = \rho_{00}$ & 0.022 & 0.0011 & 0.144\\
\hline
 $T_0$ = $T_{00}$ \newline $\rho_0 = 0.1\,\, \rho_{00}$ & 0.22 & 0.011 & 0.0144\\
\hline
 $T_0$ = $T_{00}$ \newline $\rho_0 = 2.1 \,\,\rho_{00}$ & 0.0108 & 0.0005 & 0.303\\
\hline
\end{tabular}
\caption{The values of different dimensionless ratios for the range of temperature and densities taken (where $T_{00} = 1$ MK and $\rho_{00} = 1.67 \times 10^{-12}$\,  kg\, ${\text m}^{-3}$)}
\end{table}
Figure 1a shows the dependence of density phase shift on $\rho^\prime$ for a constant temperature $T^\prime =1 $, while Figure 1b shows its dependence on $T^\prime$ for a constant density $\rho^\prime =1$. The green-solid curves are obtained by substituting the numerical solutions of wavenumber $k$ into Equation 49. The black-dashed curves are the analytical solutions in the weak damping approximation calculated using Equation 71 for comparison. Figure 1c shows the temperature phase shift similarly obtained by substituting the numerical solutions of wavenumber $k$ into Equation 55 (green-solid) and it is compared with the weak-damping approximation (black-dotted) obtained using Equation 72. Clearly the weak-damping approximation does not hold in the regime of low equilibrium density (since $d \propto \frac{1}{\rho_0}$, which implies $d \gg 1$) as indicated from the deviation of the black-dashed curves with respect to green-solid ones. However, in most of the considered range of equilibrium temperatures, the weak-damping approximation can hold well. The top and middle panels of Figure 1 match closely with the results of \citet{2009A&A...494..339O} and thus confirm their calculations. Figures 1e and 1f show the phase difference (in degrees) between temperature and density perturbations. Note that wherever we show the curves with respect to equilibrium density [$\rho^\prime$], the equilibrium temperature is kept constant at $T^\prime =1$ and the curves with respect to equilibrium temperature are always plotted keeping a constant equilibrium density $\rho^\prime =1$. In Figures 1e and 1f, the green-solid curves are calculated by substituting the wavenumber $k$ obtained from the dispersion relation into Equation 60. At this point we would like to mention that the values of phase difference obtained from Equation 60 exactly match with those obtained from Equation 67. The fact that both values exactly match is a check on the consistency of our analysis since Equation 60 is derived from continuity and momentum equations while Equation 67 from the energy equation. Equation 67 gives the general explicit expression for phase difference between density and temperature of propagating slow waves when we consider only thermal conductivity. The black-dashed curves of bottom panels in Figure 1 show the phase difference obtained for the case of the weak-damping approximation given by Equation 68. The observed value of $\Delta \phi \approx 50^\circ$ measured for a loop with $T_0$ = 1.7\,MK and $\rho_0 = 3.4 \times 10^{-12}$\,kg  ${\text m}^{-3}$ \citep{2011ApJ...727L..32V} cannot be explained by thermal conductivity alone, since for such a loop $d = 0.024$ and from Equation 68, which is a good approximation as suggested from Figure 1, we have $\Delta \phi \approx 14^\circ$ which is 3.6 times smaller.
This suggests that additional effects should be taken into account and are important for explaining the observed phase difference.

In the next section we will include viscous dissipation and study the phase shifts under the joint effect of thermal conductivity and viscosity.
\newline
\begin{figure}[h!]   
   \centerline{\hspace*{0.03\textwidth}
               \includegraphics[width=0.63\textwidth,clip=]{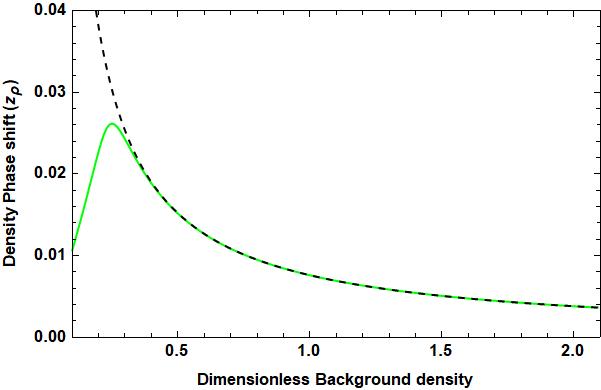} 
               \hspace*{0.06\textwidth}
               \includegraphics[width=0.63\textwidth,clip=]{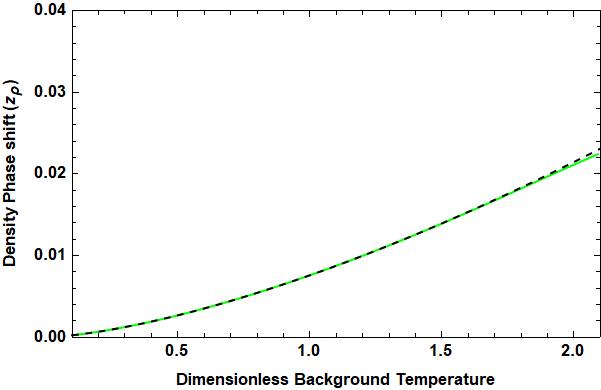}
              }
     \vspace{-0.35\textwidth}   
     \centerline{\Large \bf     
      \hspace{-0.25 \textwidth}  \color{black}{\footnotesize{(a)}}
      \hspace{0.66\textwidth}  \color{black}{\footnotesize{(b)}}
         \hfill}
     \vspace{0.315\textwidth}    
   \centerline{\hspace*{0.015\textwidth}
               \includegraphics[width=0.63\textwidth,clip=]{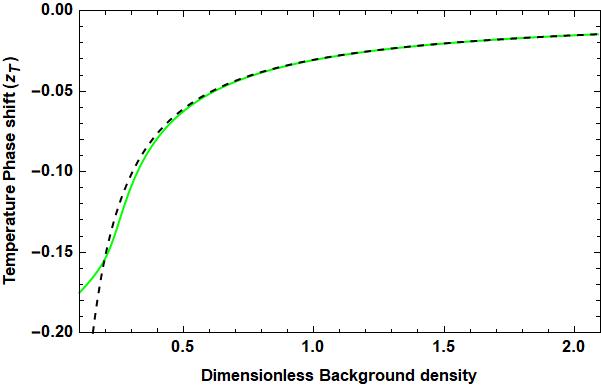}
               \hspace*{0.06\textwidth}
               \includegraphics[width=0.63\textwidth,clip=]{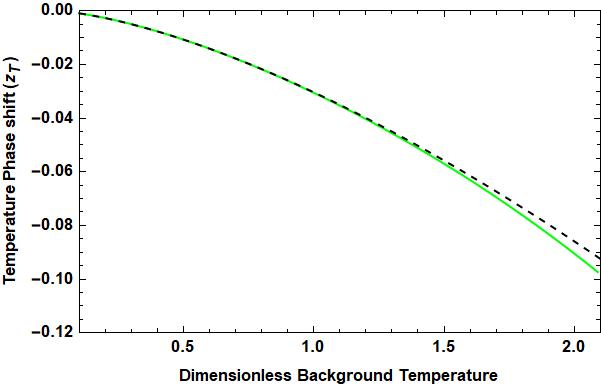}
              }
     \vspace{-0.35\textwidth}   
     \centerline{\Large \bf     
      \hspace{-0.25 \textwidth} \color{black}{\footnotesize{(c)}}
      \hspace{0.66\textwidth}  \color{black}{\footnotesize{(d)}}
         \hfill}
     \vspace{0.315\textwidth}    
   \centerline{\hspace*{0.015\textwidth}
               \includegraphics[width=0.63\textwidth,clip=]{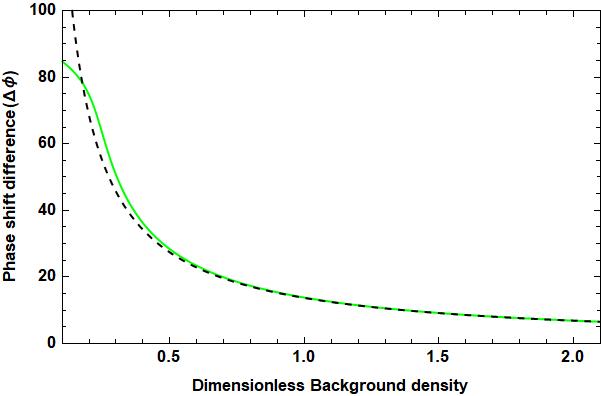}
               \hspace*{0.06\textwidth}
               \includegraphics[width=0.63\textwidth,clip=]{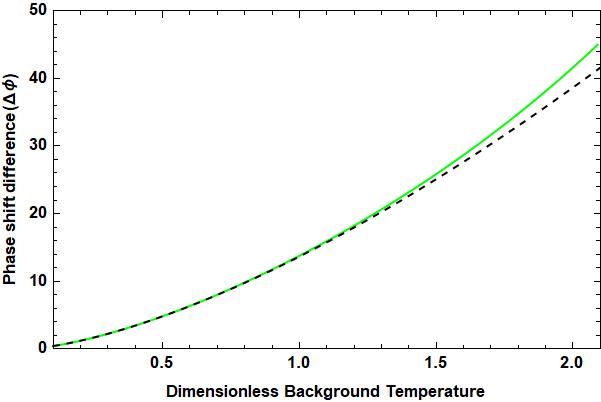}
              }
     \vspace{-0.35\textwidth}   
     \centerline{\Large \bf     
      \hspace{-0.25 \textwidth} \color{black}{\footnotesize{(e)}}
      \hspace{0.66\textwidth}  \color{black}{\footnotesize{(f)}}
         \hfill}
     \vspace{0.365\textwidth}    
\caption{In top panels $z_\rho$ is the phase shift of perturbed density relative to velocity. In the middle panels $z_T$ is the phase shift of perturbed temperature relative to velocity. In bottom panels $\Delta \phi$ is the phase difference between the perturbed density and temperature. In each panel the green-solid curves are obtained using the numerical solutions of the dispersion relation (Equation 56) while the black-dotted curves are the analytical approximations obtained under the assumption of weak damping. Note that all the panels of the left column are for a constant temperature of $T_0 = 1$\,MK while the panels of the right column are for a constant density of $\rho_{0} = 1.67 \times 10^{-12}$\,kg\,${\text m}^{-3}$ } 
 \label{F-4panels}
 \end{figure}
}
\subsection{Analysis of the Effect of Thermal Conductivity and Viscosity}
 {
Considering the joint effect of thermal conductivity and viscosity ($r=0$) we simplify Equation 43 as below
\begin{equation}
    Ak^4 + Bk^2 + C=0,
\end{equation}
where
\begin{equation}
    A = -\gamma d({\text i} + \gamma \omega e),
\end{equation}
\begin{equation}
    B = {\text i}\omega ({\text i} + \gamma \omega e) + {\text i}\gamma^2 \omega^2 d - \omega (\gamma-1),
\end{equation}
\begin{equation}
    C = \gamma \omega^3.
\end{equation}
In order to understand the role of compressive viscosity on phase shifts, we compare the solutions obtained from Equation 79 with the case when only thermal conductivity was considered.
In Figure 2 we similarly plot the phase shifts of density and temperature as was described for Figure 1. Considering the combined effect of thermal conductivity and viscosity, the red-dashed curves of Figures 2a and 2b show the variation of the density phase shift with respect to the dimensionless equilibrium density and temperature respectively. These curves are obtained by solving the dispersion relation Equation 79 and substituting the numerical solutions in Equation 49. Similar curves in Figures 2c and 2d show the same variations of temperature phase shift. We compare all the curves obtained under the joint effect of the thermal conductivity and viscosity with the similar ones when only thermal conductivity is considered as described in Section 3.1 (green-solid). The top panels of Figure 2 show that viscosity has a small but visible effect on density phase shifts in the regime of lower equilibrium density (Figure 2a) or higher equilibrium temperature (Figure 2b). This could be explained by the fact that viscous ratio $e$ has larger values in this regime. However, Figures 2c and 2d clearly show that viscosity has not much effect on temperature phase shift even for low equilibrium density or high equilibrium temperature. In Figures 2e and 2f, the red-dashed curves represent the phase difference (in degrees) when both viscosity and thermal conductivity were considered. Here also the phase difference is obtained by substituting the solutions of dispersion relation Equation 79 into Equation 60. The green-solid curves of the bottom panels show the same corresponding variations under the effect of thermal conductivity only (cf. Section 3.1). Since both curves (red-dashed and green-solid) are seen to almost superimposed on each other for the entire range of equilibrium densities and temperatures considered, we can infer that the effect of viscosity on phase difference is negligible. Thus viscosity plays nearly no role in explaining the observed phase difference between the density and temperature perturbations.

In the next section, we also consider radiative losses and a constant background heating into our model.
 \begin{figure}[h!]   
   \centerline{\hspace*{0.03\textwidth}
               \includegraphics[width=0.63\textwidth,clip=]{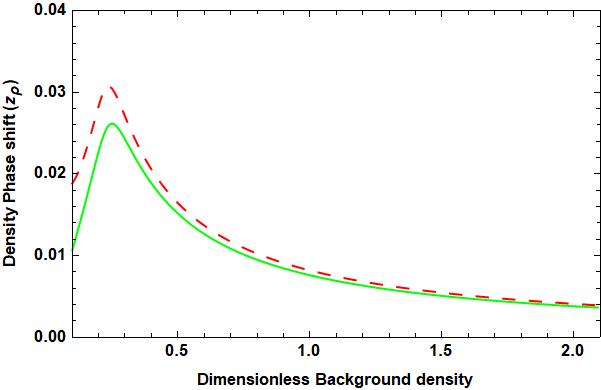}
               \hspace*{0.06\textwidth}
               \includegraphics[width=0.63\textwidth,clip=]{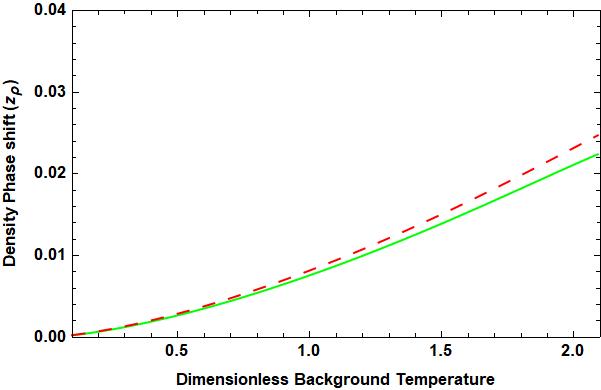}
              }
     \vspace{-0.35\textwidth}   
     \centerline{\Large \bf     
      \hspace{-0.25 \textwidth}  \color{black}{\footnotesize{(a)}}
      \hspace{0.66\textwidth}  \color{black}{\footnotesize{(b)}}
         \hfill}
     \vspace{0.315\textwidth}    
   \centerline{\hspace*{0.015\textwidth}
               \includegraphics[width=0.63\textwidth,clip=]{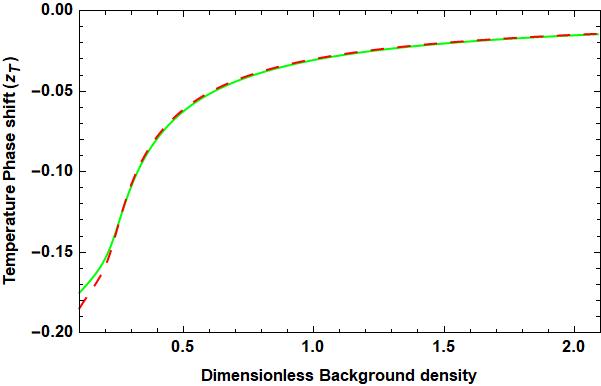}
               \hspace*{0.06\textwidth}
               \includegraphics[width=0.63\textwidth,clip=]{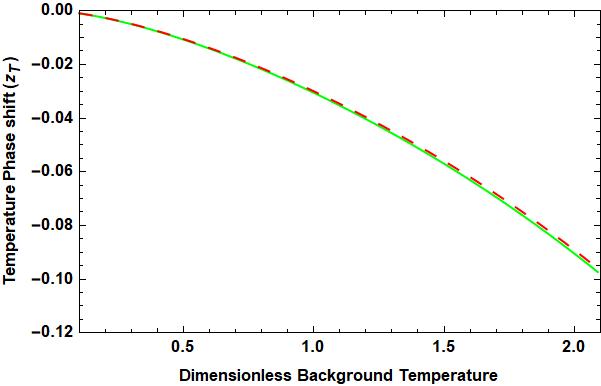}
              }
     \vspace{-0.35\textwidth}   
     \centerline{\Large \bf     
      \hspace{-0.25 \textwidth} \color{black}{\footnotesize{(c)}}
      \hspace{0.66\textwidth}  \color{black}{\footnotesize{(d)}}
         \hfill}
     \vspace{0.315\textwidth}    
   \centerline{\hspace*{0.015\textwidth}
               \includegraphics[width=0.63\textwidth,clip=]{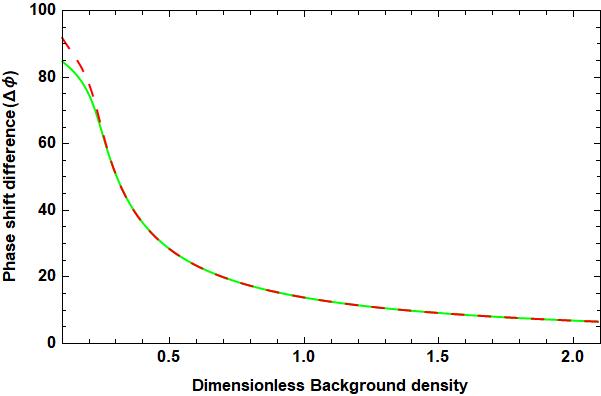}
               \hspace*{0.06\textwidth}
               \includegraphics[width=0.63\textwidth,clip=]{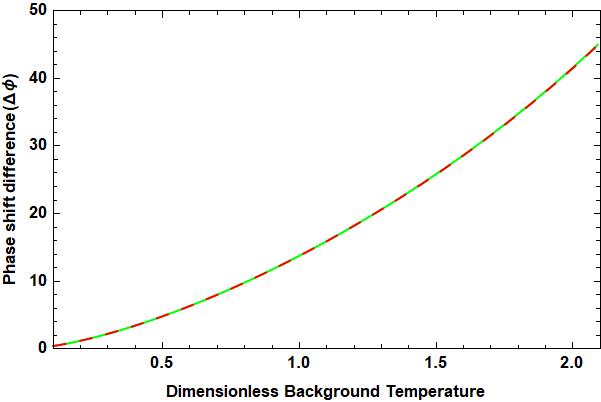}
              }
     \vspace{-0.35\textwidth}   
     \centerline{\Large \bf     
      \hspace{-0.25 \textwidth} \color{black}{\footnotesize{(e)}}
      \hspace{0.66\textwidth}  \color{black}{\footnotesize{(f)}}
         \hfill}
     \vspace{0.375\textwidth}    
     
\caption{In top panels $z_\rho$ is the phase shift of perturbed density relative to velocity. In the middle panels $z_T$ is the phase shift of perturbed temperature relative to velocity. In bottom panels $\Delta \phi$ is the phase difference between the perturbed density and temperature. In each panel the red-dashed curves are obtained under the joint effect of thermal conductivity and viscosity while the green-solid curves are the ones when only thermal conductivity is considered. Note that all the panels of the left column are for a constant temperature of $T_0 = 1$\,MK while the panels of the right column are for a constant density of $\rho_{0} = 1.67 \times 10^{-12}$\,kg\,${\text m}^{-3}$ } 
 \label{F-4panels}
 \end{figure}
}
\subsection{Analysis of the Joint Effect of Thermal Conductivity, Viscosity and Radiative Losses with Constant Background Heating per Unit Mass}
{
We consider the joint effect of radiative losses along with thermal conductivity, viscosity, and a constant background heating ($a=b=0$). A constant background heating balances the initial radiative losses so as to maintain a thermal equilibrium,
\begin{equation*}
    H(\rho,T) = \chi\rho_0T_0^\alpha = \text{constant}
\end{equation*}
We have the following simplified expression for the dispersion relation
\begin{equation}
    Ak^4 + Bk^2 + C=0,
\end{equation}
where
\begin{equation}
    A = -\gamma d({\text i} + \gamma \omega e),
\end{equation}
\begin{equation}
    B = ({\text i}\omega - \alpha \gamma r) ({\text i} + \gamma \omega e) + {\text i}\gamma^2 \omega^2 d - \omega (\gamma-1) + {\text i}\gamma r,
\end{equation}
\begin{equation}
    C = -{\text i}\gamma \omega^2({\text i}\omega - \alpha \gamma r).
\end{equation}
Figure 3 shows similar curves as to those in Figure 2, however, here the orange-dot--dashed curves are obtained considering the joint effect of thermal conductivity, viscosity, and radiative losses with a constant background heating. The orange-dot--dashed curves here are obtained in a similar manner using the solutions of dispersion relation Equation 83. The red-dashed curves in each panel of Figure 3 represent the same corresponding variations when thermal conductivity and viscosity are considered (cf. Section 3.2). Figure 3a  shows that the density phase shift is significantly increased by the addition of radiative losses and a constant background heating for higher equilibrium densities, while the effect is negligible for low equilibrium densities considered. Figure 3b shows that the effect of radiative losses is significant in the regime of low temperatures as compared to that for higher temperatures. The middle row of Figure 3 shows that the addition of radiative losses has only a weak effect on the variation of the temperature phase shift with density (Figure 3c), while its effect is significant for the lower temperature (Figure 3d). In the bottom panels of Figure 3, the phase difference is significantly increased by the addition of radiative losses and constant heating for lower equilibrium temperatures. We find that the radiative loss has significant effect on the phase shifts and the phase difference for lower temperature. This is consistent with the fact that radiative ratio $r$ has larger values in this regime (cf. Table 1).  However, for the coronal loop with $T^\prime =1$ and $\rho^\prime =1$ the inclusion of radiative losses and constant background heating has not much effect. We calculate that for the loop with $T_0 =1.7$\,MK and $\rho_0 = 3.4 \times 10^{-12}$\,kg $\text{m}^{-3}$ the phase difference $\Delta \phi \approx 17^\circ$ which cannot explain the observations of \citet{2011ApJ...727L..32V}. 

In the next section we consider the effect of heating--cooling imbalance and discuss its effect on phase shifts.
 \begin{figure}[h!]   
   \centerline{\hspace*{0.03\textwidth}
               \includegraphics[width=0.63\textwidth,clip=]{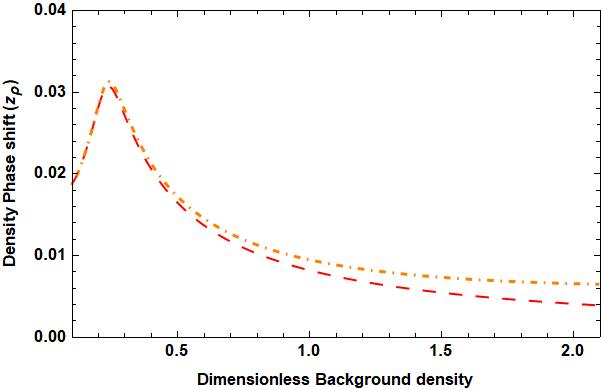}
               \hspace*{0.06\textwidth}
               \includegraphics[width=0.63\textwidth,clip=]{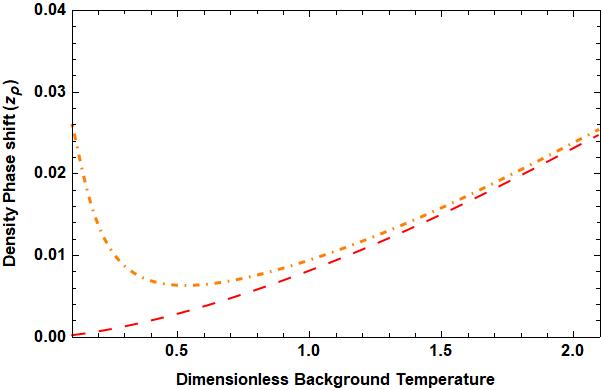}
              }
     \vspace{-0.35\textwidth}   
     \centerline{\Large \bf     
      \hspace{-0.25 \textwidth}  \color{black}{\footnotesize{(a)}}
      \hspace{0.66\textwidth}  \color{black}{\footnotesize{(b)}}
         \hfill}
     \vspace{0.315\textwidth}    
   \centerline{\hspace*{0.015\textwidth}
               \includegraphics[width=0.63\textwidth,clip=]{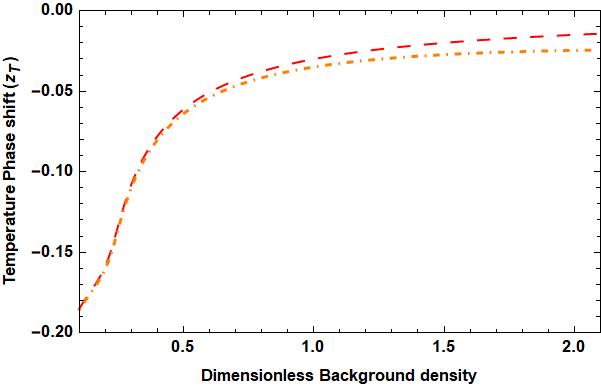}
               \hspace*{0.06\textwidth}
               \includegraphics[width=0.63\textwidth,clip=]{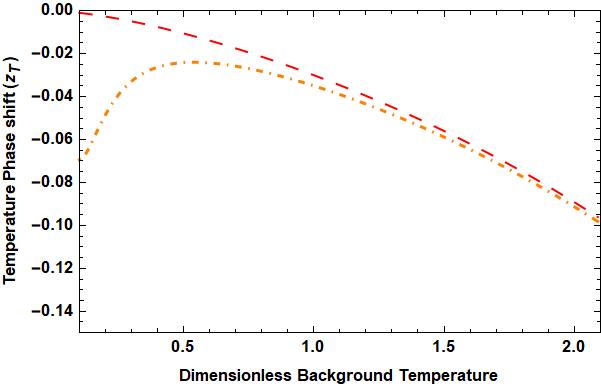}
              }
     \vspace{-0.35\textwidth}   
     \centerline{\Large \bf     
      \hspace{-0.25 \textwidth} \color{black}{\footnotesize{(c)}}
      \hspace{0.66\textwidth}  \color{black}{\footnotesize{(d)}}
         \hfill}
     \vspace{0.315\textwidth}    
   \centerline{\hspace*{0.015\textwidth}
               \includegraphics[width=0.63\textwidth,clip=]{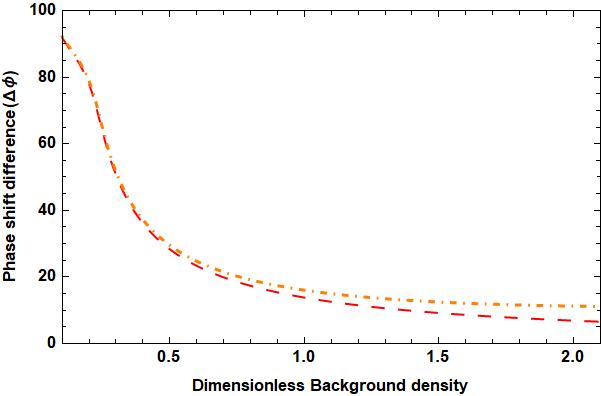}
               \hspace*{0.06\textwidth}
               \includegraphics[width=0.63\textwidth,clip=]{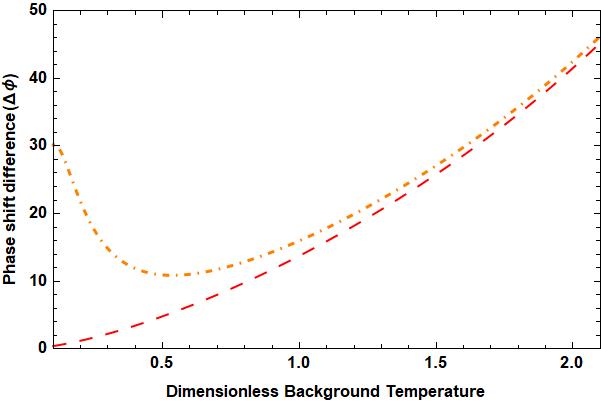}
              }
     \vspace{-0.35\textwidth}   
     \centerline{\Large \bf     
      \hspace{-0.25 \textwidth} \color{black}{\footnotesize{(e)}}
      \hspace{0.66\textwidth}  \color{black}{\footnotesize{(f)}}
         \hfill}
     \vspace{0.365\textwidth}    
     
\caption{In top panels $z_\rho$ is the phase shift of perturbed density relative to velocity. In the middle panels $z_T$ is the phase shift of perturbed temperature relative to velocity. In bottom panels $\Delta \phi$ is the phase difference between the perturbed density and temperature. In each panel the orange-dot--dashed curves are obtained under the joint effect of thermal conductivity, viscosity and radiative losses with constant heating while the red-dashed curves are the ones when only thermal conductivity and viscosity is considered. Note that all the panels of the left column are for a constant temperature of $T_0 = 1$\,MK while the panels of the right column are for a constant density of $\rho_{0} = 1.67 \times 10^{-12}$\,kg\,${\text m}^{-3}$ } 
 \label{F-4panels}
 \end{figure}
}
\subsection{Analysis of the Joint Effect of Thermal Conductivity, Viscosity and Radiative Losses with Heating--Cooling Imbalance}
{
Considering the joint effect of thermal conductivity, viscosity and radiative losses with heating--cooling imbalance we have the expression
\begin{equation}
    Ak^4 + Bk^2 + C=0,
\end{equation}
where 
\begin{equation}
    A = -\gamma d({\text i} + \gamma \omega e),
\end{equation}
\begin{multline}
    B = ({\text i} + \gamma \omega e)({\text i}\omega + \gamma r(b - \alpha)) + {\text i}\gamma^2\omega^2d - \omega(\gamma-1) - {\text i}\gamma r(a-1),
\end{multline}
\begin{equation}
    C = -{\text i}\gamma \omega^2({\text i}\omega + \gamma r(b - \alpha)).
\end{equation}
We consider a heating function of the form
\begin{equation}
    H(\rho,T) \propto \rho^{-\frac{1}{2}}T^{-3}.
\end{equation}
The choice of this heating function is a particular instance considered, and throughout our analysis we shall consider this heating function in the case of heating--cooling imbalance.
Initially the radiative losses are balanced by the heating function (cf. Equation 12) which maintains thermal equilibrium. As the slow MHD waves propagate in the plasma medium, they cause local perturbations in the background-plasma parameters such as density, temperature, and pressure, which eventually leads to a wave-induced imbalance between the radiative losses and assumed coronal heating leading to a thermal misbalance. This physical scenario affects the propagating slow waves by modifying the dispersion relation (cf. Equation 87).

Further it is clarified that in the present model we do not consider any effect of radiative losses on the evolution of the background plasma which is assumed isothermal and homogeneous (cf. Section 2.1). This consideration is quite reasonable since the radiative-cooling timescale $\left[\tau_{\rm rad} = \frac{\tau}{r}\right]$ \citep{2003A&A...408..755D} is quite long in comparison to the wave period. For the temperature range of $T_0$ = 1\,--\,2\,MK at $\rho_0 = 1.67 \times 10^{-12}$\, kg $\text{m}^{-3}$, $\tau_{\rm rad}$ is in the range of $\approx$ 2000\,--\,6000 seconds, which is an order of magnitude larger as compared to the wave period of $\tau \approx 300$ seconds. The heating--cooling imbalance in the present context only refers to the wave-induced thermal misbalance between the plasma heating and cooling processes. 

In the top and middle panels of Figure 4 we plot the density and temperature phase shifts (blue-dashed curve) considering the combined effect of thermal conductivity, viscosity, radiative loss with heating--cooling imbalance. In each panel we have compared the blue-dashed curves with the case when constant heating was considered as described in Section 3.3 (orange-dot--dashed curves). Heating--cooling imbalance is found to significantly increase the density phase shift at higher equilibrium densities (Figure 4a) and lower equilibrium temperatures (Figure 4b). Similarly, in the middle panels the temperature phase shift is reduced significantly by the addition of heating--cooling imbalance in the similar regime of higher densities or lower temperatures. In the bottom panels, we observe that the phase difference increases drastically when the equilibrium temperature is reduced to 0.1\,MK (Figure 4e). In conclusion, the effect of heating--cooling imbalance is found to be greater in loops of higher densities and lower temperatures.

We find that for the case of $T_0 = 1.7$\,MK and $\rho_0 = 3.4 \times 10^{-12}$\,kg $\text{m}^{-3}$ the inclusion of heating--cooling imbalance gives $\Delta \phi \approx 24^\circ$.
Thus the joint effect of thermal conductivity, viscosity, radiative losses, and heating--cooling imbalance with the considered heating function is not very successful in explaining the observed phase difference of typical coronal loops \citep{2011ApJ...727L..32V, 2015ApJ...811L..13W, 2018ApJ....860...107, 2018ApJ...868..149K}. In the next section we will discuss the polytropic index obtained from the linear MHD model and determine the conditions under which the observations can be matched with the theoretical analysis.
\begin{figure}[h!]   
   \centerline{\hspace*{0.03\textwidth}
               \includegraphics[width=0.63\textwidth,clip=]{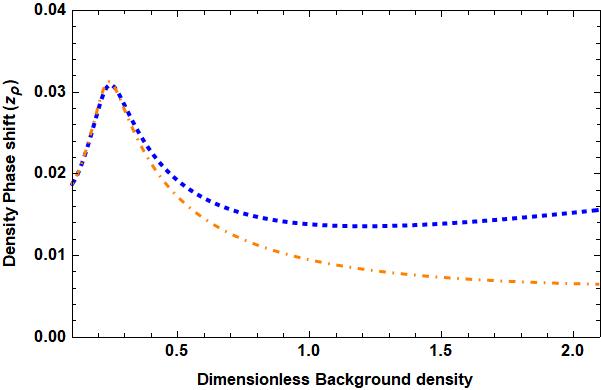}
               \hspace*{0.06\textwidth}
               \includegraphics[width=0.63\textwidth,clip=]{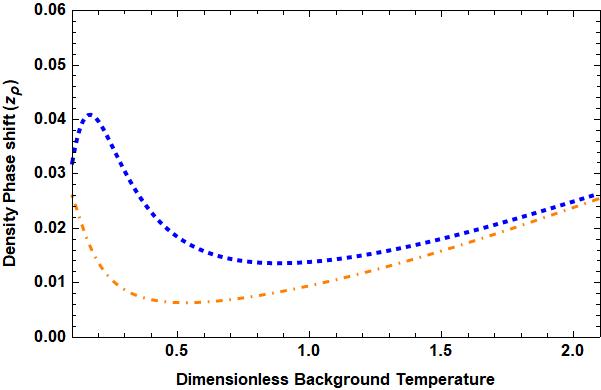}
              }
     \vspace{-0.35\textwidth}   
     \centerline{\Large \bf     
      \hspace{-0.25 \textwidth}  \color{black}{\footnotesize{(a)}}
      \hspace{0.66\textwidth}  \color{black}{\footnotesize{(b)}}
         \hfill}
     \vspace{0.315\textwidth}    
   \centerline{\hspace*{0.015\textwidth}
               \includegraphics[width=0.63\textwidth,clip=]{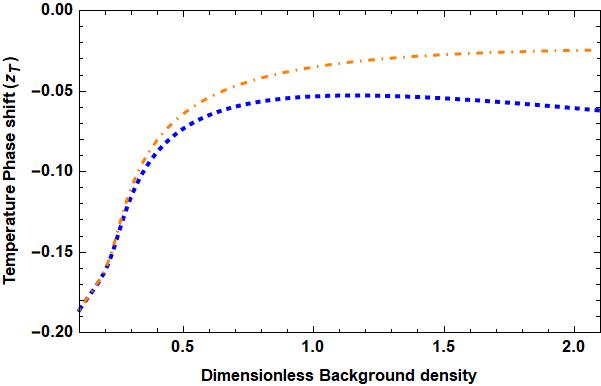}
               \hspace*{0.06\textwidth}
               \includegraphics[width=0.63\textwidth,clip=]{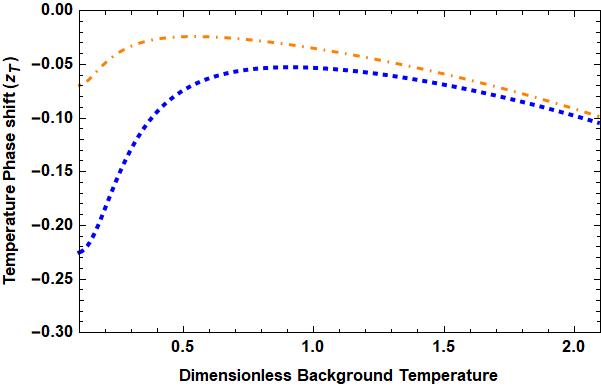}
              }
     \vspace{-0.35\textwidth}   
     \centerline{\Large \bf     
      \hspace{-0.25 \textwidth} \color{black}{\footnotesize{(c)}}
      \hspace{0.66\textwidth}  \color{black}{\footnotesize{(d)}}
         \hfill}
     \vspace{0.315\textwidth}    
   \centerline{\hspace*{0.015\textwidth}
               \includegraphics[width=0.63\textwidth,clip=]{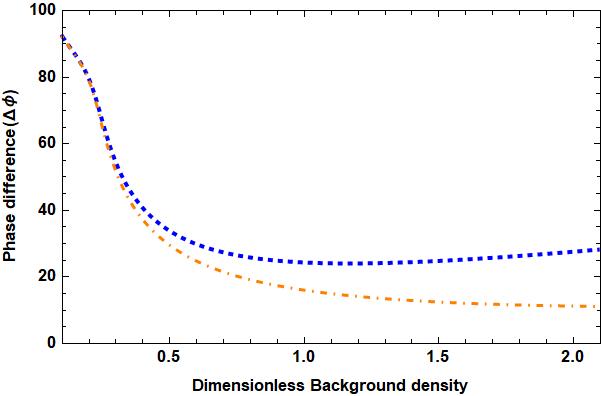}
               \hspace*{0.06\textwidth}
               \includegraphics[width=0.63\textwidth,clip=]{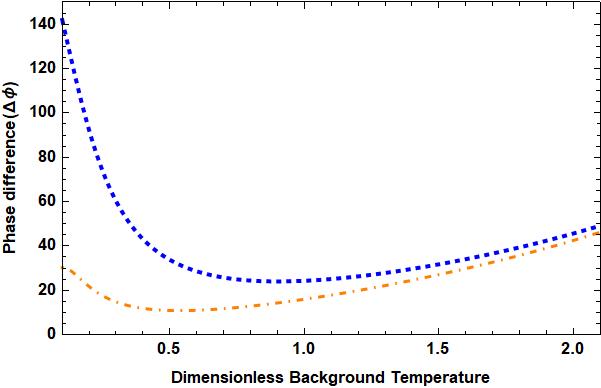}
              }
     \vspace{-0.35\textwidth}   
     \centerline{\Large \bf     
      \hspace{-0.25 \textwidth} \color{black}{\footnotesize{(e)}}
      \hspace{0.66\textwidth}  \color{black}{\footnotesize{(f)}}
         \hfill}
     \vspace{0.355\textwidth}    
     
\caption{In top panels $z_\rho$ is the phase shift of perturbed density relative to velocity. In the middle panels $z_T$ is the phase shift of perturbed temperature relative to velocity. In bottom panels $\Delta \phi$ is the phase difference between the perturbed density and temperature. In each panel the blue-dashed curves are obtained under the joint effect of thermal conductivity, viscosity, and radiative losses with heating--cooling imbalance while the orange-dot--dashed curves are the ones when constant heating is considered. Note that all the panels of the left column are for a constant temperature of $T_0 = 1$\,MK while the panels of the right column are for a constant density of $\rho_{0} = 1.67 \times 10^{-12}$\,kg\,${\text m}^{-3}$ }  
 \label{F-4panels}
 \end{figure}
}
\subsection{Polytropic Index}
{
The polytropic index is a very important quantity in coronal seismology and has been estimated from the observed data on slow magnetoaccostic waves in many previous works \citep{2011ApJ...727L..32V, 2015ApJ...811L..13W, 2018ApJ....860...107, 2018ApJ...868..149K}. We derive here the general theoretical expression for the polytropic index using linear MHD taking into account all of the effects of thermal conductivity, viscosity, radiative losses, and heating--cooling imbalance.\newline
Considering the linearized energy equation,
\begin{equation}
    \frac{\partial T_1}{\partial t} = -(\gamma-1)\frac{\partial v_1}{\partial z} + \gamma d  \left(  \frac{\partial^2 T_1}{\partial z^2} \right) - \gamma r(\alpha T_1 + \rho_1) + \gamma r(bT_1 + a\rho_1),
\end{equation}
from the mass conservation equation we can write
\begin{equation}
    \frac{\partial T_1}{\partial t} = (\gamma-1)\frac{\partial \rho_1}{\partial t} + \gamma d  \left(  \frac{\partial^2 T_1}{\partial z^2} \right) + \gamma r(b-\alpha)T_1 + \gamma r(a-1)\rho_1.
\end{equation}
Since we have
\begin{equation}
    \rho_1 = \hat{\rho}_1 \,\,{\text e}^{{\text i}(kz-\omega t)},
\end{equation}
\begin{equation}
    T_1 = \hat{T}_1\,\, {\text e}^{{\text i}(kz-\omega t - \Delta \phi)},
\end{equation}
Thus substituting in the energy equation we get
\begin{equation}
    {\text e}^{-{\text i}\Delta \phi}(\gamma dk^2 - \gamma r(b - \alpha) - {\text i}\omega )\hat{T}_1 = (\gamma r(a-1) - {\text i}\omega (\gamma-1))\hat{\rho}_1.  
\end{equation}
Writing the real and imaginary parts of the above equation separately we can get
\begin{equation}
    \left[ \cos \Delta \phi (2\gamma dk_{\rm r}k_{\rm i} - \omega) - \sin \Delta \phi (\gamma d(k_{\rm r}^2 - k_{\rm i}^2) - \gamma r(b-\alpha))  \right]\hat{T}_1 = -\omega (\gamma -1)\hat{\rho}_1,
\end{equation}
\begin{equation}
    \left[ \sin \Delta \phi (2\gamma dk_{\rm r}k_{\rm i} - \omega) +  \cos \Delta \phi (\gamma d(k_{\rm r}^2 - k_{\rm i}^2) - \gamma r(b-\alpha))  \right]\hat{T}_1 = \gamma r(a-1)\hat{\rho}_1.
\end{equation}
multiplying Equation 97 by $\cos \Delta \phi$ and Equation 98 by $\sin \Delta \phi$ then adding we get
\begin{equation}
    \hat{T}_1 =\frac{1}{\omega - 2\gamma dk_{\rm r}k_{\rm i}}\left( \omega(\gamma -1)\cos \Delta \phi-\gamma r(a-1)\sin \Delta \phi  \right)\hat{\rho}_1 \equiv (\gamma_{\rm eff} -1)\hat{\rho}_1,
\end{equation}
where $\gamma_{\rm eff}$ is defined based on the polytropic assumption \citep{2015ApJ...811L..13W, 2018ApJ....860...107}; thus we have
\begin{equation}
    \gamma_{\rm eff} -1 = \frac{1}{\omega - 2\gamma dk_{\rm r}k_{\rm i}}\left( \omega(\gamma -1)\cos \Delta \phi-\gamma r(a-1)\sin \Delta \phi  \right).
\end{equation}
In the case with only thermal conductivity (i.e. $r=0$) and the weak-damping assumption ($k_{\rm i}=0$) we recover the original expression from \citet{2018ApJ....860...107}, \citet{2018ApJ...868..149K}, \citet{2011ApJ...727L..32V}:
\begin{equation}
    \gamma_{\rm eff} -1 = (\gamma -1)\cos \Delta \phi.
\end{equation}
In Figure 5 we plot the polytropic index with respect to the dimensionless background density and temperature considering only the effect of thermal conductivity. We have plotted $\gamma_{\rm eff}$ (green-solid) by substituting the numerical solution of dispersion relation Equation 56 into Equation 100 with $r=0$ and compared it with the weak-damping approximation (black-dashed) given by Equation 101. As expected, the weak damping does not hold for low equilibrium densities or high equilibrium temperatures ( with $d \gg 1$). We have also plotted the $\gamma_{\rm eff}$ (red-dashed) considering the joint effect of thermal conductivity and compressive viscosity by substituting the solutions of dispersion relation Equation 79 into Equation 100 with $r=0$. We find that the overall effect of compressive viscosity on $\gamma_{\rm eff}$ is negligible for the entire range of temperatures and densities considered. It is interesting to note that in most of the considered range of densities and temperatures, the polytropic index is closer to the theoretical value of $1.66$. However, at low densities when $\rho_0 < 0.3\rho_{00}$ (Figure 5a) the polytropic index decreases drastically and becomes close to the inferred value of $1.10$ at $\rho_0 = 0.1\rho_{00}$. This suggests that to achieve the observed polytropic index the thermal ratio [$d$] needs to be much larger than its classical value (cf. Table 1 and discussion later on). 

In Figure 6 we plot the polytropic index with respect to the dimensionless background density and temperature considering the joint effect of thermal conductivity, viscosity, and radiative losses. The orange-dot--dashed curves in the top panels of Figure 6 show the variation of the polytropic index in the presence of constant heating while the blue-dashed curves in bottom panels show the same variations with heating--cooling imbalance. We have also compared all the plots with the case when only thermal conductivity was considered (green-solid). Overall, from Figure 6 we can see that the inclusion of radiative losses with the constant heating or heating--cooling imbalance does not have a significant effect on the values of the polytropic index. From Figures 6a and 6c we see that the effect of radiative losses  on the polytropic index is less affected by changes in equilibrium density. Figure 6b shows that the presence of radiative losses with constant heating increases the polytropic index from its classical value only when the equilibrium temperature is low. When heating--cooling imbalance is considered, the polytropic index is slightly reduced from its classical value at lower equilibrium temperatures (Figure 6d). 

We find that the joint effect of thermal conductivity, viscosity, radiative losses, and heating--cooling imbalance with the chosen heating function ($a=-0.5$ and $b=-3$) cannot explain the observed polytropic index of $\gamma_{\rm eff} = 1.10 \pm 0.02$, \citep{2011ApJ...727L..32V}. In order to bring the polytropic index to its observed value we expand our range of temperature and densities further. In the top panels of Figure 7, we show the variation of the phase difference with equilibrium density (Figure 7a) and equilibrium temperature (Figure 7b). Keep in mind that all the curves with respect to equilibrium density are plotted keeping equilibrium temperature constant at $T_0 = 1$\,MK, and the curves with respect to equilibrium temperature are obtained by considering a constant density of $\rho_0 = 1.67 \times 10^{-12}$\,kg ${\text m}^{-3}$. In the middle panels of Figure 7 we show the variation of the polytropic index with equilibrium density and temperature. Note that here we have considered temperatures up to 10\,MK and densities up to 30 times larger than $1.67 \times 10^{-12}$\,kg ${\text m}^{-3}$. We observe that the polytropic index decreases to its reported value \citep[]{2018ApJ...868..149K, 2011ApJ...727L..32V} when the equilibrium density is quite low with $\rho_0 \approx 10^{-13}$\,kg ${\text m}^{-3}$ (Figure 7c) or the equilibrium temperature is much higher at $T_0 \approx 6.5$\,MK (Figure 7d) compared to the measured coronal values. This again suggests an expected thermal ratio much larger than that calculated in Table 1. 

In order to determine an estimate of the thermal and radiative ratio that can account for the observed polytropic index, we plot the polytropic index with respect to both ratios in the bottom panels of Figure 7. Figure 7e is obtained by keeping the background temperature fixed at $T_0= 1$\,MK  and varying the equilibrium density, while Figure 7f is obtained by changing the equilibrium temperature keeping the equilibrium density constant ($\rho_0 = 1.67 \times 10^{-12}$\,kg ${\text m}^{-3}$). For the polytropic index to reach a value close to observations, the thermal ratio should be sufficiently large ($d \approx 0.3$) about 14 times larger compared to the classical value for $T_0=T_{00}$ and $\rho_0 = \rho_{00}$ given in Table 1. While the radiative ratio should be sufficiently small ($r \approx 0.008$), about 18 times smaller  in comparison to the the range of values given in Table 1. However, it is very unlikely for such a drastic reduction in the radiative effects to happen with the very precise atomic measurements of radiative losses in the solar abundance (CHIANTI) .

From the expressions for thermal and radiative ratios (cf. Equations 76 and 78) it is clearly seen that a regime of high equilibrium temperature or low equilibrium density leads to higher values of $d$ along with a simultaneous reduction of $r$. Thus a natural question arises whether it is the combined effect of anomalously high thermal ratio and low radiative ratio that is bringing the polytropic index to its inferred value, or whether a high thermal ratio can alone explain the measurements. In order to investigate further we artificially increased $d$ by an order of magnitude while keeping the radiative ratio [$r$] at its classical value (cf. Table 1). Interestingly we found that when the thermal ratio was increased, the phase difference for the loop with $T_{0} = 1$\,MK and $\rho_{00} = 1.67 \times 10^{-12}$\,kg ${\text m}^{-3}$ became $\Delta \phi = 90.5^\circ$ while $\gamma_{\rm eff} = 1.16$. Since reducing the radiative ratio is not reasonable, as mentioned above, this suggests that it is likely the anomalous behaviour leading to strong thermal conductivity that can explain the observations of \citet{2011ApJ...727L..32V}. The inferref polytropic index can be achieved when thermal conductivity is highly efficient compared to the expected value from the classical Spitzer theory in the condition considered.

Further, we also consider the effect of different forms of heating functions by varying the power indices $a$ and $b$ from $-5$ to $5$. In Figure 8 we plot the phase difference and polytropic index for background temperature of $T_0 = 1$\,MK and background density of $\rho_0 = 1.67 \times 10^{-12}$\,kg $\text{m}^{-3}$. Figure 8a shows the variation of phase difference, and we can observe that it is highly dependent on the form of heating function, which may lead to a better explanation of observed phase difference ($\Delta \phi \approx 50^\circ$). The top-right corner of Figure 8a shows that there is a phase reversal between temperature and density perturbations at higher values of $a$ and $b$. The polytropic index remains close to its classical value for the majority of the range of $a$ and $b$ considered (cf. Figure 8b). We can clearly see that considering different forms of heating functions does not successfully explain the inferred polytropic index, and this further supports the idea that the enhanced thermal conductivity can better explain the observations.   

\begin{figure}[t!]   
   \centerline{\hspace*{0.03\textwidth}
               \includegraphics[width=0.63\textwidth,clip=]{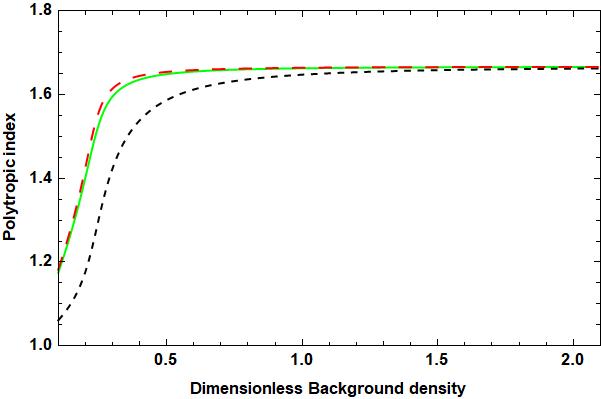}
               \hspace*{0.06\textwidth}
               \includegraphics[width=0.63\textwidth,clip=]{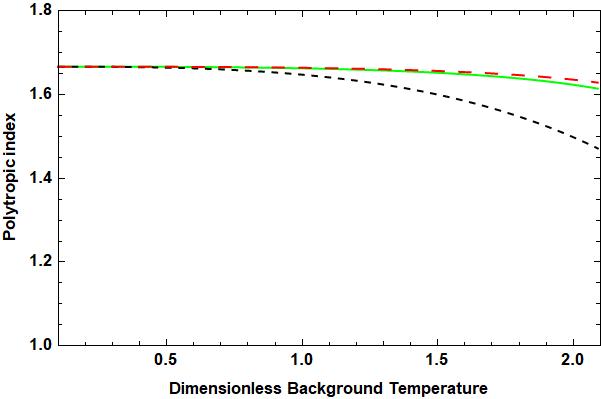}
              }
     \vspace{-0.35\textwidth}   
     \centerline{\Large \bf     
      \hspace{-0.25 \textwidth}  \color{black}{\footnotesize{(a)}}
      \hspace{0.66\textwidth}  \color{black}{\footnotesize{(b)}}
         \hfill}
     \vspace{0.315\textwidth}    
\caption{The left panel shows the variation of polytropic index with background density at constant temperature of $T_0 = 1$\,MK and right panel shows the same variation with respect to background temperature at constant density of $\rho_0 = 1.67 \times 10^{-12}$\,kg ${\text m}^{-3}$. In each panel the green-solid curves are obtained using the numerical solutions of the dispersion relation (Equation 56) when only thermal conductivity is considered while the black-dashed curves are the corresponding analytical approximations obtained under the assumption of weak damping. The red-dashed curves are obtained under the joint effect of thermal conductivity and compressive viscosity.} 
 \label{F-4panels}
 \end{figure}

\begin{figure}[h!]   

   \centerline{\hspace*{0.015\textwidth}
               \includegraphics[width=0.63\textwidth,clip=]{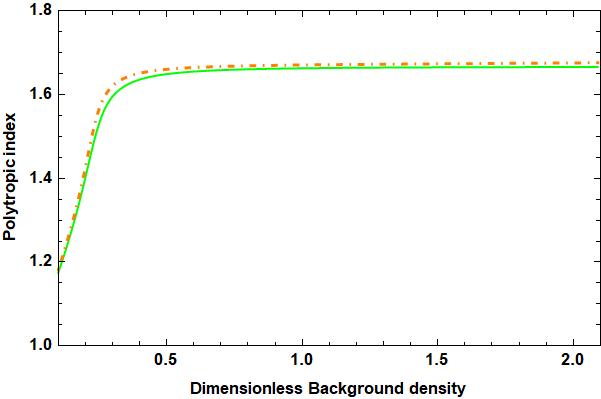}
               \hspace*{0.06\textwidth}
               \includegraphics[width=0.63\textwidth,clip=]{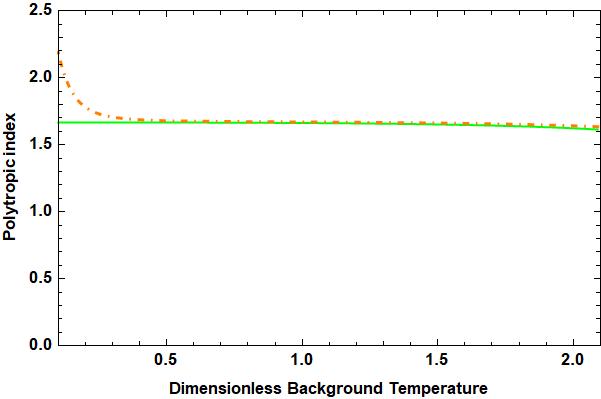}
              }
     \vspace{-0.35\textwidth}   
     \centerline{\Large \bf     
      \hspace{-0.25 \textwidth} \color{black}{\footnotesize{(a)}}
      \hspace{0.66\textwidth}  \color{black}{\footnotesize{(b)}}
         \hfill}
     \vspace{0.315\textwidth}    
   \centerline{\hspace*{0.015\textwidth}
               \includegraphics[width=0.63\textwidth,clip=]{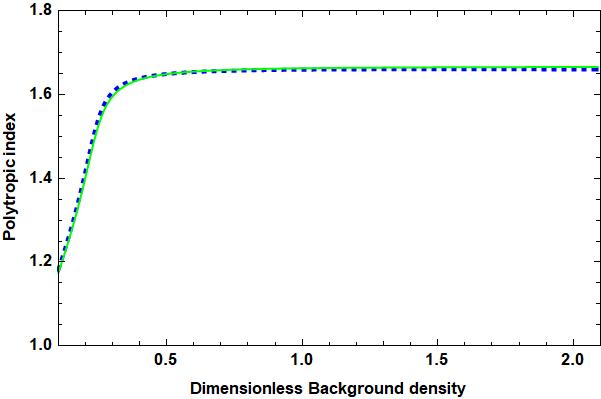}
               \hspace*{0.06\textwidth}
               \includegraphics[width=0.63\textwidth,clip=]{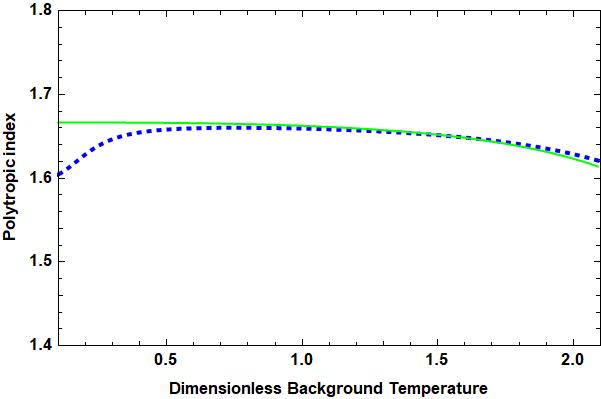}
              }
     \vspace{-0.35\textwidth}   
     \centerline{\Large \bf     
      \hspace{-0.25 \textwidth} \color{black}{\footnotesize{(c)}}
      \hspace{0.66\textwidth}  \color{black}{\footnotesize{(d)}}
         \hfill}
     \vspace{0.315\textwidth}    

\caption{The left panels show the variation of polytropic index with background density at constant temperature of $T_0 = 1$\,MK and right panels show the same variation with respect to background temperature at constant density of $\rho_0 = 1.67 \times 10^{-12}$\,kg ${\text m}^{-3}$. In top panels the orange-dot--dashed curves are obtained under the joint effect of thermal conductivity, compressive viscosity, and radiative losses with constant heating while in the bottom panels the blue-dashed curves are for the case with heating-cooling imbalance. In each panel the green-solid curves are obtained for the case when only thermal conductivty is considered.} 
 \label{F-4panels}
 \end{figure}
 \begin{figure}[h!]   

   \centerline{\hspace*{0.015\textwidth}
               \includegraphics[width=0.63\textwidth,clip=]{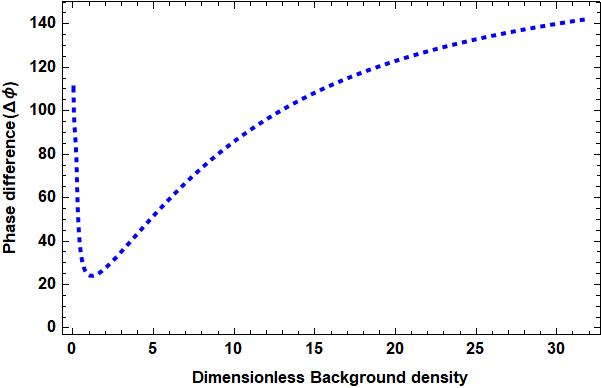}
               \hspace*{0.06\textwidth}
               \includegraphics[width=0.63\textwidth,clip=]{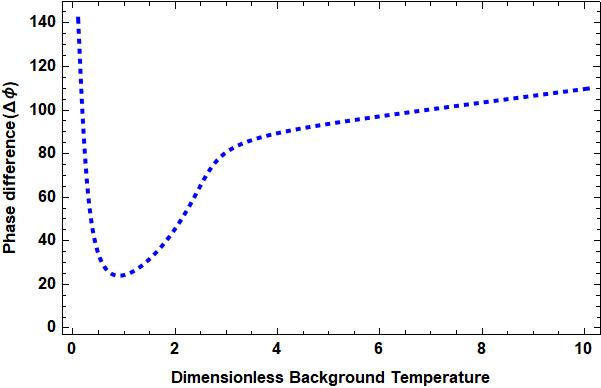}
              }
     \vspace{-0.35\textwidth}   
     \centerline{\Large \bf     
      \hspace{-0.25 \textwidth} \color{black}{\footnotesize{(a)}}
      \hspace{0.66\textwidth}  \color{black}{\footnotesize{(b)}}
         \hfill}
     \vspace{0.325\textwidth}    
   \centerline{\hspace*{0.015\textwidth}
               \includegraphics[width=0.63\textwidth,clip=]{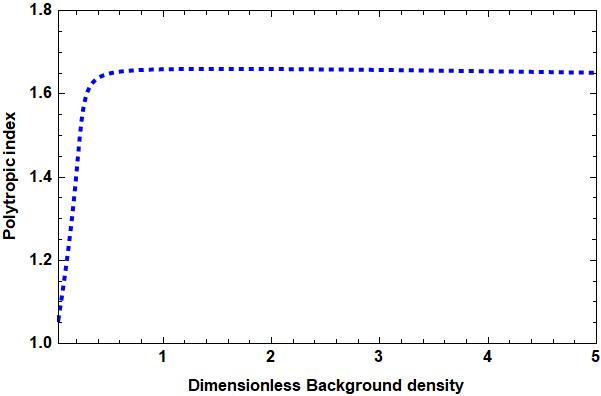}
               \hspace*{0.06\textwidth}
               \includegraphics[width=0.63\textwidth,clip=]{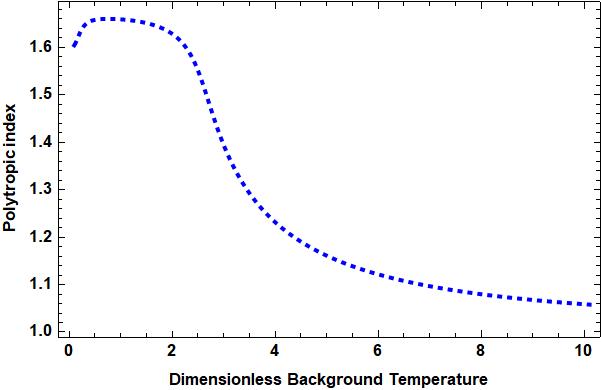}
              }
     \vspace{-0.35\textwidth}   
     \centerline{\Large \bf     
      \hspace{-0.25 \textwidth} \color{black}{\footnotesize{(c)}}
      \hspace{0.66\textwidth}  \color{black}{\footnotesize{(d)}}
         \hfill}
     \vspace{0.325\textwidth}    
   \centerline{\hspace*{0.015\textwidth}
               \includegraphics[width=0.63\textwidth,clip=]{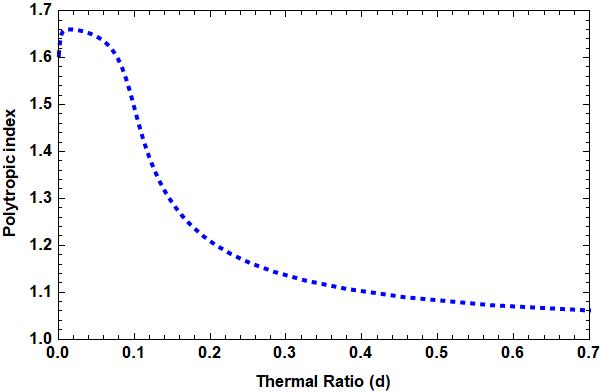}
               \hspace*{0.06\textwidth}
               \includegraphics[width=0.63\textwidth,clip=]{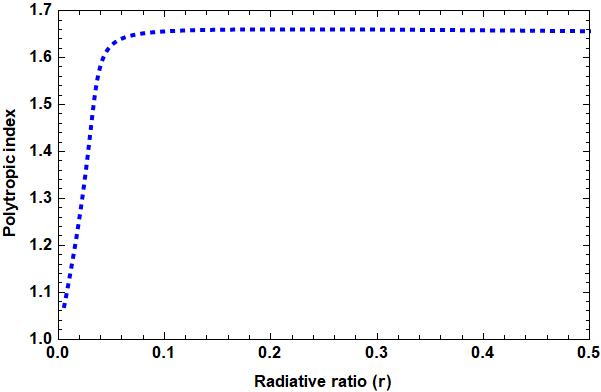}
              }
     \vspace{-0.35\textwidth}   
     \centerline{\Large \bf     
      \hspace{-0.25 \textwidth} \color{black}{\footnotesize{(e)}}
      \hspace{0.66\textwidth}  \color{black}{\footnotesize{(f)}}
         \hfill}
     \vspace{0.365\textwidth}    

\caption{In top panels $\Delta \phi$ is the phase difference between perturbed density and temperature. In the middle panels the variation of polytropic index with background density and temperature is shown. In the bottom panels the variation of polytropic index with thermal ratio [$d$] and radiative ratio [$r$] is shown. The blue-dashed curves are obtained under the joint effect of thermal conductivity, viscosity, and radiative losses with heating--cooling imbalance. Note that all the panels of the left column are for a constant temperature of $T_0 = 1$\,MK while the panels of the right column have a constant density of $\rho_{0} = 1.67 \times 10^{-12}$\,kg ${\text m}^{-3}$ } 
 \label{F-4panels}
 \end{figure}
 \begin{figure}[t!]   
   \centerline{\hspace*{0.03\textwidth}
               \includegraphics[width=0.63\textwidth,clip=]{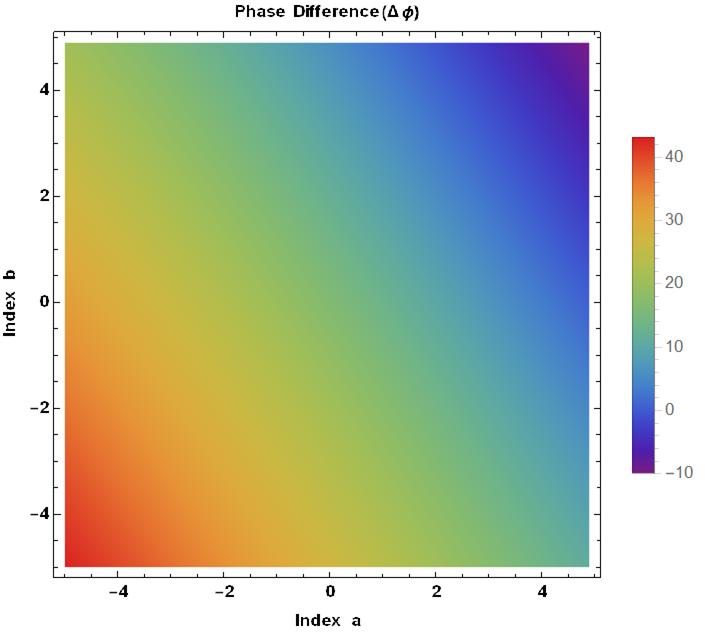}
               \hspace*{0.06\textwidth}
               \includegraphics[width=0.63\textwidth,clip=]{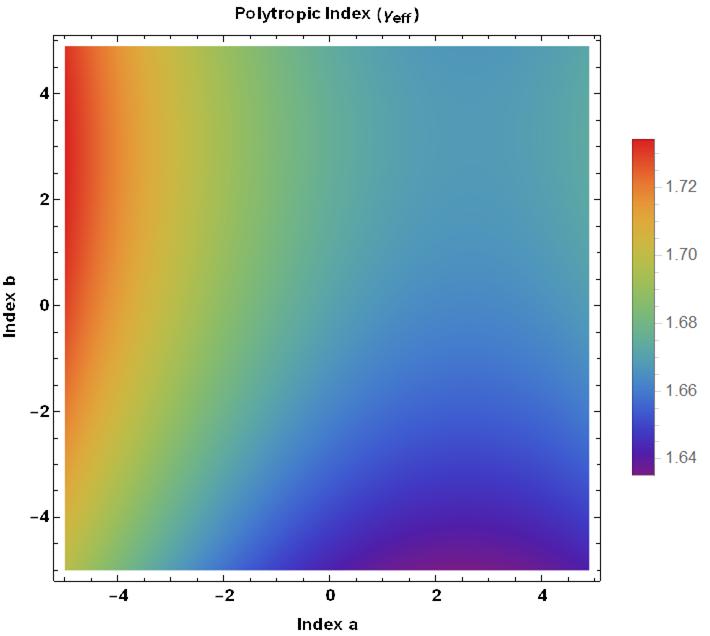}
              }
     \vspace{-0.35\textwidth}   
     \centerline{\Large \bf     
      \hspace{-0.25 \textwidth}  \color{black}{\footnotesize{(a)}}
      \hspace{0.66\textwidth}  \color{black}{\footnotesize{(b)}}
         \hfill}
     \vspace{0.315\textwidth}    
\caption{The left panel shows the variation of phase difference with power indices $a$ and $b$ while the right panel shows similar variation of polytropic index. Both panels are plotted for a constant background temperature of $T_0 =1$\,MK and background density of $\rho_0 = 1.67 \times 10^{-12}$\,kg ${\text m}^{-3}$.} 
 \label{F-4panels}
 \end{figure}
}
\section{Discussion and Conclusion}
{
We have developed a comprehensive linear model for the propagating slow MHD waves  to explain the observed phase shifts of density and temperature perturbations in  warm ($T$ = 1\,--\,2\,MK) coronal loops.  We estimate the density phase and temperature phase with respect to velocity perturbations  of the slow waves, and their variations with respect to the background densities and temperatures within the coronal loops. We have also estimated the phase difference between density and temperature perturbations and their dependence on the background density and temperature. We have chosen a range of the coronal loops in our study with densities and temperatures commonly observed in the warm corona for the propagating slow waves. However we have ignored the effect of gravitational stratification in our analysis, which does have some influence on warm $T$ = 1\,--\,2\,MK loops \citep{2009A&A...494..339O} and is beyond the scope of the current study. We derived the general dispersion relation taking into account the joint effect of thermal conductivity, viscosity, radiative losses, and heating--cooling imbalance, thus invoking the most comprehensive physical scenario of the non-adiabatic coronal-loop plasma. In Section 3.1,  we studied the effect of thermal conductivity on density and temperature phase shifts and our results closely match those of the previous work of \citet{2009A&A...494..339O}. However, thermal conductivity alone could not explain the observed phase difference as reported by \citet{2011ApJ...727L..32V}, and we infer that other physical effects in our MHD model may be important for the better understanding of the physics of propagating slow waves. Therefore, step-by-step, we have incorporated the viscous damping, radiative losses, and heating--cooling imbalance into our model, which were subsequently discussed in Sections 3.2, 3.3, and 3.4. For heating--cooling imbalance, we considered a particular heating function with $a=-0.5$ and $b=-3$, and that is related to the damped oscillatory mode of the slow mode waves \citep{2020arXiv201114519P}. Note that the heating--cooling imbalance in the present work does not refer to the effect of radiative losses on the evolution of background plasma but rather implies the wave-induced thermal misbalance in the plasma medium (cf. Section 3.4). The effect of viscosity was found to be negligible on the phase shifts and phase difference of density and temperature perturbations. The joint effect of radiative losses, and heating--cooling imbalance increased the density phase shift and reduced the temperature phase shift in comparison to the case when only thermal conductivity was considered, but it was significant mostly for high equilibrium (or background) density or low equilibrium (or background) temperature of the coronal loop. However, the combined effect of thermal conductivity, viscosity, radiative losses, and heating--cooling imbalance in our model cannot explain the observed phase difference
($\Delta \phi \approx 50^{\circ}$) for the typical coronal loops as measured with {\it Hinode}/EIS by \citet{2011ApJ...727L..32V} and \citet{2018ApJ...868..149K}.

In Section 3.5, we derived the general expression for the polytropic index [$\gamma_{\rm eff}$] considering the joint effect of thermal conductivity, compressive viscosity, and radiative losses with heating--cooling imbalance (cf. Equation 100). It is mentioned here that \citet{2019Aip...26...8} have also studied the effective adiabatic index under the effect of heating--cooling imbalance, however in the present work we follow a different definition from theirs and provide a more generalized expression. We find that the effect of compressive viscosity on polytropic index is much weaker compared to the other mechanisms considered. Although for most of the loop parameters considered, the polytropic index is close to the classical value $1.66$, however, it becomes close to a value of 1.2 when the loop density is reduced by an order of magnitude. This motivated us to expand our range of equilibrium temperatures and densities to look for the regime where the polytropic index lies close to its inferred value ($\gamma_{\rm eff} \approx 1.1$) in typical coronal loops \citep{2011ApJ...727L..32V,2018ApJ...868..149K}. We found that when the equilibrium density is reduced by an order of magnitude and/or the equilibrium temperature is increased by an order of magnitude, then the expected polytropic index can lie in the range of the observed values as reported by \citet{2011ApJ...727L..32V},  however the loops of such low densities or high temperatures are not consistent with direct measurements from {\it Hinode}/EIS and SDO/AIA. This instead implies the possible anomalously higher thermal conductivity compared to the classical value. In this regime of loops, for polytropic index to have a value $\approx 1.1$ \citep{2011ApJ...727L..32V}, the thermal ratio [$d$] should be $\approx$14 times larger, and the radiative ratio [$r$] approximately 18 times smaller in comparison to their respective values in typical coronal loops (cf. Table 1). 

We artificially increased the thermal ratio [$d$] by an order of magnitude while keeping the density and temperature in the range for typical coronal conditions. We found that this increase of thermal ratio [$d$] can match the observed values of $\gamma_{\rm eff}$. Since the enhancement of $d$ cannot be explained by uncertainties in measurements of loop temperature [$T_{0}$] and density [$\rho_{0}$] while the artificial change in radiative ratio is unreasonable when $T_0$ and $\rho_0$ are given, it suggests a significant enhancement of the classical thermal-conduction coefficient. It appears that the observations of \citet{2011ApJ...727L..32V} and \citet{2018ApJ...868..149K} belong to a region ofthe  solar corona with anomalously large thermal conductivity, and the exact reasons for this behaviour need to be investigated in future studies. This conclusion is based on the linear MHD model developed in the present work and suggests that thermal conductivity should be greater than that expected from the classical theory. However, the enhanced thermal conductivity alone still cannot consistently explain the observed phase difference. This suggests that some other effects need to be considered, such as different forms of heating function.

We also considered different forms of heating functions for a fixed background temperature and density of 1\,MK and $1.67 \times 10^{-12}$\,kg\,m$^{-3}$ respectively (cf. Figure 8). We found that for most of the values of $a$ and $b$ the polytropic index remains close to or is higher than its classical value. However, the phase difference is strongly dependent on the form of the heating function and may lead to a better explanation of the observed phase difference ($\Delta \phi \approx 50^\circ$). 

To the best of our knowledge, this work is the first attempt to study phase shifts of propagating slow waves using a theoretical linear MHD model incorporating the effect of thermal conductivity along with compressive viscosity, radiative losses, and heating--cooling imbalance. The present work would be useful for future studies and the interpretation of observations of propagating slow waves since the large parametric range considered in the study (cf. Figure 7) would allow the potential works to compare with the observations and determine the key factors affecting the phase shifts. This new model adds a comprehensive view and a diagnostic capability for coronal seismology using propagating slow waves.

Finally, we would like to mention that in the current study we have assumed a fixed radiative function ($\alpha = -0.5$) for the entire range of temperatures and densities. The chosen radiative function is quite a reasonable approximation in the range of temperatures (0.1\,MK to 2\,MK) considered in our present work. We fitted the more realistic piece-wise radiative function of \citet{2008ApJ...682..1351} in a range of 0.1\,MK to 100\,MK and found that the fitted power index [$\alpha$] is -0.58, which is quite close to the assumed index of -0.5 (Priest, 2014). The distinct difference is observed only for  temperatures less than 0.1 MK, which is beyond the consideration of the temperatures 0.1\,MK\,--\,2\,MK in our study where the radiative function used is reasonably close to the piece-wise function given by \citet{2008ApJ...682..1351}

Although the fixed power-law radiative function chosen in the present work is a good approximation, however it is limited in capturing the local features of the more realistic functions given by \citet{2008ApJ...682..1351} and CHIANTI atomic database. Since heating--cooling imbalance is shown to be highly dependent on these detailed features of the radiative function such as the local gradients, the present analysis and results may be limited to the approximation considered. An important and useful extension of the present study would be to include more realistic radiative function in the presented linear MHD model for a better understanding of the role of thermal misbalance on phase shifts of propagating slow waves in the solar corona.

}
\begin{acks}
We thank the reviewer for their constructive comments that improved our manuscript.
A. Prasad thanks IIT (BHU) for the computational facility, and A.K. Srivastava acknowledges the support of UKIERI (Indo-UK) research grant for the present research. 
The work of T.J. Wang was supported by NASA grants 80NSSC18K1131 and 80NSSC18K0668 as well as the NASA Cooperative Agreement NNG11PL10A to CUA.
A.K. Srivastava also acknowledges the ISSI-BJ regarding the science team project on ``Oscillatory Processes in Solar and Stellar Coronae''.
\end{acks}\newline\newline
{\footnotesize {\bf Disclosure of Potential Conflicts of Interest:}\newline The authors declare that there are no conflicts of interest.}

\end{article} 

\end{document}